\documentclass[submission, Phys]{SciPost}

\usepackage[utf8]{inputenc} 
\usepackage[T1]{fontenc} 	
\usepackage[english]{babel} 

\usepackage[bitstream-charter]{mathdesign}
\usepackage{geometry} 		
\usepackage{amsmath} 		
\usepackage{mathtools} 		
\usepackage{float} 			
\usepackage{graphicx} 		
\usepackage{tabularx} 		
\usepackage{booktabs} 		
\usepackage{xltabular} 		
\usepackage{pdflscape} 		
\usepackage{color, xcolor} 	
\usepackage{pdfpages} 		
\usepackage{extarrows} 		
\usepackage{multirow} 		
\usepackage{multicol} 		
\usepackage{enumitem} 		
\usepackage{xspace} 		
\usepackage{stackrel} 		
\usepackage{tikz} 			
\usepackage{braket} 		
\usepackage{bm} 			
\usepackage{tensor} 		
\usepackage{slashed} 		
\usepackage{siunitx} 		
\usepackage{lastpage} 		
\usepackage{cite} 			
\usepackage[normalem]{ulem} 
\usepackage{fontawesome} 	
\usepackage{tocloft} 		
\usepackage{titlesec} 		
\usepackage{doi} 			
\usepackage{hyperref} 		
\usepackage[nameinlink, capitalize]{cleveref} 	
\usepackage[nottoc, notlot, notlof]{tocbibind} 	
\usepackage[ruled, vlined]{algorithm2e} 		
\usepackage{makecell}
\usepackage[makeroom]{cancel}
\usepackage{feynmf}

\makeatletter
\def\BState{\State\hskip-\ALG@thistlm}
\makeatother

\makeatletter
\@ifundefined{pdfoutput}{}{\DeclareGraphicsRule{*}{mps}{*}{}}
\makeatother

\makeatletter
\DeclareRobustCommand*{\bfseries}{%
   \not@math@alphabet\bfseries\mathbf
   \fontseries\bfdefault\selectfont
   \boldmath
}
\makeatother

\hypersetup{
	pdftitle={MadAgents},
	pdfauthor={Plehn et al.},
	colorlinks=true, 			
	linkcolor={red!50!black}, 	
	citecolor={blue!50!black}, 	
	urlcolor={blue!80!black} 	
} 

\DeclareSymbolFont{usualmathcal}{OMS}{cmsy}{m}{n}
\DeclareSymbolFontAlphabet{\mathcal}{usualmathcal}

\SetArgSty{textnormal}
\SetKwComment{Comment}{{\small\#}~}{}
\SetCommentSty{mycommfont}

\setitemize{itemsep=0pt, parsep=0pt} 				
\setenumerate{itemsep=0pt, parsep=0pt} 				
\newlist{todolist}{itemize}{2}
\setlist[todolist]{label=$\square$}
\usepackage{pifont}
\setitemize{itemsep=2pt,topsep=2pt,parsep=0pt,partopsep=0pt,leftmargin=*}
\setenumerate{itemsep=0pt,topsep=2pt,parsep=0pt,partopsep=0pt,labelindent=3pt,leftmargin=*}
\usepackage{amsmath}
 
\usepackage{amsthm} 		
\theoremstyle{definition}

\usepackage{listings}
\usepackage{listingsutf8}

\usepackage[verbatim,cachedir=m-cache]{minted}
\usepackage[most]{tcolorbox}
\tcbuselibrary{minted,skins,breakable}
\usepackage{fvextra}
 
\newcommand{\appsubsection}[1]{
    \stepcounter{subsection}
    \subsection*{\Alph{section}.\arabic{subsection}\hspace{1em}{#1}}
}

\definecolor{sci_light_blue}{HTML}{4869b4}
\definecolor{sci_light_tes}{HTML}{a08aaa}
\definecolor{sci_light_gre}{HTML}{3E5F58}
\definecolor{red_cb}{HTML}{e41a1c}
\definecolor{blue_cb}{HTML}{377eb8}
\definecolor{green_cb}{HTML}{4daf4a}
\definecolor{purple_cb}{HTML}{984ea3}
\definecolor{orange_cb}{HTML}{ff7f00}

\definecolor{EmeraldGreen}{HTML}{1ea78d}
\definecolor{EnglishRed}{HTML}{b02427}
\hypersetup{colorlinks=true,urlcolor=EmeraldGreen,citecolor=EmeraldGreen,linkcolor=EnglishRed}

\newcommand{\eg}{\text{e.g.}\;}

\newcommand\one{\leavevmode\hbox{\small1\normalsize\kern-.33em1}}

\newcommand{\ma}{\textsc{MadAgents}\xspace}
\newcommand{\herwig}{\textsc{Herwig}\xspace}

\newcommand{\mg}{\textsc{MadGraph}\xspace}
\newcommand{\pythia}{\textsc{Pythia}\xspace}
\newcommand{\fastjet}{\textsc{FastJet}\xspace}

\newcommand{\delphes}{\textsc{Delphes}\xspace}

\newcommand{\sherpa}{\textsc{Sherpa}\xspace}

\newcommand{\madnis}{\textsc{MadNIS}\xspace}

\newcommand{\lhapdf}{\textsc{Lhapdf~6}\xspace}
\newcommand{\python}{\textsc{Python}\xspace}

\renewcommand{\root}{\textsc{Root}\xspace}

\newcommand{\arXiv}[2][]{%
	\ifthenelse{\equal{#1}{}}%
	{\href{http://arxiv.org/abs/#2}{arXiv:#2}}%
	{\href{http://arxiv.org/abs/#2}{arXiv:#2~[#1]}}}

\def\slashchar#1{\setbox0=\hbox{$#1$}           
   \dimen0=\wd0                                 
   \setbox1=\hbox{/} \dimen1=\wd1               
   \ifdim\dimen0>\dimen1                        
      \rlap{\hbox to \dimen0{\hfil/\hfil}}      
      #1                                        
   \else                                        
      \rlap{\hbox to \dimen1{\hfil$#1$\hfil}}   
      /                                         
   \fi}

\newcommand{\tikznode}[2]{%
\ifmmode%
\tikz[remember picture,baseline=(#1.base),inner sep=0pt] \node (#1) {$#2$};%
\else
\tikz[remember picture,baseline=(#1.base),inner sep=0pt] \node (#1) {#2};%
\fi}

\def\mathswitchr#1{\relax\ifmmode{\mathrm{#1}}\else$\mathrm{#1}$\xspace\fi}
\def\mathswitch#1{\relax\ifmmode#1\else$#1$\xspace\fi}

\makeatletter
\newcommand{\indentafterblock}{%
  \futurelet\pb@next\pb@after
}
\newcommand{\pb@after}{%
  \ifx\pb@next\par
  \else
    \ifhmode\par\fi
    \noindent\ignorespaces
  \fi
}
\makeatother

\newtcblisting{bashblock}{
  listing engine=minted,
  minted language=bash,
  breakable,
  enhanced,
  colback=black!2,
  colframe=black!20,
  arc=2pt,
  boxrule=0.45pt,
  left=4pt,right=4pt,top=2pt,bottom=2pt,
  boxsep=1pt,
  listing only,
  before skip=4pt, after skip=4pt,
  after=\indentafterblock,
  minted options={
    breaklines,
    breaksymbolleft=,
    fontsize=\footnotesize,
    autogobble,
    tabsize=2
  }
}

\tcbuselibrary{skins,breakable,listings}

\tcbset{
  promptblock/base/.style={
    breakable,
    enhanced,
    arc=2pt,
    boxrule=0.45pt,
    left=4pt,right=12pt,top=4pt,bottom=3pt,
    boxsep=1pt,
    title={#1},
    title filled,
    coltitle=black,
    fontupper=\footnotesize,
    fonttitle=\ttfamily\small\bfseries,
    toptitle=1.5pt,bottomtitle=1.5pt,
    lefttitle=4pt,righttitle=4pt,
    before skip=4pt, after skip=4pt,
    after=\indentafterblock
  }
}

\tcbset{
  promptblock/plan/.style={
    colback=sci_light_tes!5,
    colframe=sci_light_tes!50,
    colbacktitle=sci_light_tes!50
  },
  promptblock/user/.style={
    colback=sci_light_gre!5,
    colframe=sci_light_gre!50,
    colbacktitle=sci_light_gre!50 
},
  promptblock/final/.style={
    colback=sci_light_blue!5,
    colframe=sci_light_blue!50,
    colbacktitle=sci_light_blue!50
  }
}

\newtcolorbox{userblock}[1][]{
  promptblock/base={#1},
  promptblock/user
}

\newtcolorbox{planblock}[1][]{
  promptblock/base={#1},
  promptblock/plan
}

\newtcolorbox{finalblock}[1][]{
  promptblock/base={#1},
  promptblock/final
}

\newtcolorbox{promptblock}[1][]{
  breakable,
  enhanced,
  colback=sci_light_blue!5,
  colframe=sci_light_blue!50,
  arc=2pt,
  boxrule=0.45pt,
  left=4pt,right=12pt,top=4pt,bottom=3pt,
  boxsep=1pt,
  title={#1},
  title filled,
  colbacktitle=sci_light_blue!50,
  coltitle=black,
  fontupper=\footnotesize,
  fonttitle=\ttfamily\small\bfseries,
  toptitle=1.5pt,bottomtitle=1.5pt,
  lefttitle=4pt,righttitle=4pt,
  before skip=4pt, after skip=4pt,
  after=\indentafterblock
}

\newtcblisting{promptblocklit}[1][]{
  breakable,
  enhanced,
  colback=black!3,
  colframe=black!35,
  arc=2pt,
  boxrule=0.45pt,
  left=2pt,right=2pt,top=2pt,bottom=2pt,
  boxsep=0.5pt,
  title={#1},
  title filled,
  colbacktitle=black!20,
  coltitle=black,
  fonttitle=\ttfamily\footnotesize\bfseries,
  toptitle=1pt,bottomtitle=1pt,
  lefttitle=2pt,righttitle=2pt,
  before skip=3pt, after skip=3pt,
  listing only,
  listing engine=listings,
  listing options={
    basicstyle=\ttfamily\tiny\linespread{0.95}\selectfont,
    columns=fullflexible,
    keepspaces=true,
    showspaces=false,
    showstringspaces=false,
    breaklines=true,
    breakatwhitespace=false,
    upquote=true,
    aboveskip=0pt,
    belowskip=0pt,
    literate={—}{{---}}1 {“}{{``}}1 {”}{{''}}1 {→}{{->}}2 {–}{{--}}1 {‘}{{`}}1 {’}{{'}}1 {…}{{...}}3 {√}{{sqrt}}4
  },
  after=\indentafterblock
}

\newtcblisting{promptblocklitlarge}[1][]{
  breakable,
  enhanced,
  colback=black!3,
  colframe=black!35,
  arc=2pt,
  boxrule=0.45pt,
  left=2pt,right=2pt,top=2pt,bottom=2pt,
  boxsep=0.5pt,
  title={#1},
  title filled,
  colbacktitle=black!20,
  coltitle=black,
  fonttitle=\ttfamily\footnotesize\bfseries,
  toptitle=1pt,bottomtitle=1pt,
  lefttitle=2pt,righttitle=2pt,
  before skip=3pt, after skip=3pt,
  listing only,
  listing engine=listings,
  listing options={
    basicstyle=\ttfamily\footnotesize\linespread{0.95}\selectfont,
    columns=fullflexible,
    keepspaces=true,
    showspaces=false,
    showstringspaces=false,
    breaklines=true,
    breakatwhitespace=false,
    upquote=true,
    aboveskip=0pt,
    belowskip=0pt,
    literate={—}{{---}}1 {“}{{``}}1 {”}{{''}}1 {→}{{->}}2 {–}{{--}}1 {‘}{{`}}1 {’}{{'}}1 {…}{{...}}3 {√}{{sqrt}}4
  },
  after=\indentafterblock
}

\usepackage[fencedCode,pipeTables,texMathDollars,tightLists,frozenCache,cacheDir=md-cache]{markdown}

\AddToHook{begindocument/end}{%
  \markdownSetup{
    renderers = {
      softLineBreak = {\unskip\space},
      hardLineBreak = {\unskip\space},
    },
  }%
}

\usepackage{fvextra}
\usepackage{xurl}
\usepackage{xstring}

\usetikzlibrary{shadows.blur}

\definecolor{mdAccent}{HTML}{2563EB} 
\definecolor{mdBg}{HTML}{F8FAFC}     
\definecolor{mdFrame}{HTML}{CBD5E1}  
\definecolor{mdTitleBg}{HTML}{EFF6FF}

\tcbset{
  mdCard/.style={
    enhanced,
    breakable,
    colback=mdBg,
    colframe=mdAccent!35!mdFrame,
    arc=2.2mm,
    boxrule=0.45pt,
    left=1.8mm,right=1.8mm,top=1.3mm,bottom=1.3mm,
    boxsep=0.9mm,
    colbacktitle=mdTitleBg,
    coltitle=black,
    fonttitle=\ttfamily\small\bfseries,
    boxed title style={sharp corners, boxrule=0pt},
    toptitle=0.8mm,bottomtitle=0.8mm,
    before skip=4pt,
    after skip=4pt
  }
}

\newtcolorbox{mdcodeblock}[1][]{%
  enhanced,
  breakable,
  colback=mdBg,
  colframe=mdAccent!30!mdFrame,
  arc=2.0mm,
  boxrule=0.4pt,
  left=4.5pt,right=4.5pt,top=2.2pt,bottom=2.2pt,
  boxsep=1pt,
  #1
}

\makeatletter

\def\markdownRendererCodeSpan#1{%
  {\ttfamily\footnotesize
    \begingroup
      \def\markdownRendererUnderscore##1{\string_}%
      \def\markdownRendererBackslash##1{\string\\}%
      \def\markdownRendererLeftBrace##1{\string\{}%
      \def\markdownRendererRightBrace##1{\string\}}%
      \def\markdownRendererDollarSign##1{\string$}%
      \def\markdownRendererTilde##1{\string~}%
      \def\markdownRendererCircumflex##1{\string^}%
      \edef\md@codespan{#1}%
      \expandafter\path\expandafter{\md@codespan}%
    \endgroup
  }%
}

\makeatother

\newcommand{\markdownfile}[2][Markdown File]{%
  \begin{tcolorbox}[
    mdCard,
    title={#1},
    fontupper=\footnotesize,
    after=\indentafterblock
  ]
    \begingroup
      \renewcommand\addcontentsline[3]{}%

      \setlength{\parskip}{0pt}%
      \setlength{\parindent}{0pt}%

      \renewcommand{\section}[1]{\par\smallskip\noindent\textbf{\footnotesize ##1}\par\smallskip}
      \renewcommand{\subsection}[1]{\par\smallskip\noindent\textbf{\footnotesize ##1}\par\smallskip}
      \renewcommand{\subsubsection}[1]{\par\smallskip\noindent\textbf{\footnotesize ##1}\par\smallskip}
      \renewcommand{\paragraph}[1]{\par\smallskip\noindent\textbf{\footnotesize ##1}\par\smallskip}
      \renewcommand{\subparagraph}[1]{\par\smallskip\noindent\textbf{\footnotesize ##1}\par\smallskip}

      \setlist[itemize,1]{topsep=0pt,itemsep=0pt,parsep=0pt,partopsep=0pt}
      \setlist[itemize,2]{topsep=0pt,itemsep=0pt,parsep=0pt,partopsep=0pt}
      \setlist[enumerate,1]{topsep=0pt,itemsep=0pt,parsep=0pt,partopsep=0pt}
      \setlist[enumerate,2]{topsep=0pt,itemsep=0pt,parsep=0pt,partopsep=0pt}

      \markdownInput{#2}%
      
      \endgroup
  \end{tcolorbox}%
}

\graphicspath{{./figs/}}

\usetikzlibrary{arrows.meta,positioning,calc,fit}

\usepackage{tcolorbox}
\tcbset{promptbox/.style={
  colback=black!3,
  colframe=black!40,
  fonttitle=\bfseries,
  boxrule=0.4pt,
  arc=2pt,
  left=4pt,right=4pt,top=4pt,bottom=4pt
}}

\begin{document}

\begin{center}{\Large \textbf{
MadAgents
}}\end{center}

\begin{center}
Tilman Plehn\textsuperscript{1,2},
Daniel Schiller\textsuperscript{1}, and
Nikita Schmal\textsuperscript{1}
\end{center}

\begin{center}
{\bf 1} Institut f\"ur Theoretische Physik, Universit\"at Heidelberg, Germany\\
{\bf 2} Interdisciplinary Center for Scientific Computing (IWR), Universit\"at Heidelberg, Germany\\
\end{center}

\begin{center}
\today
\end{center}


\section*{Abstract}
{\bf 
    We uncover an effective and communicative set of agents working with \mg. Agentic installation, learning-by-doing training, and user support provide easy access to state-of-the-art simulations and accelerate LHC research. We show in detail how \ma interact with inexperienced and advanced users, support a range of simulation tasks, and analyze the results. In a second step, we illustrate how \ma automatize event generation and run an autonomous simulation campaign, starting from a pdf file of a paper. We also present successive \ma updates, including a Claude Code implementation with a self-improvement loop.
}

{
\vspace{5pt}
\noindent \url{https://github.com/MadGraphTeam/MadAgents}
}

\vspace{10pt}
\noindent\rule{\textwidth}{1pt}
\tableofcontents\thispagestyle{fancy}
\noindent\rule{\textwidth}{1pt}
\vspace{10pt}

\clearpage

\clearpage
\section{Introduction}

The challenge of the current and future LHC program is how to analyze vast amounts of complex data in terms of fundamental physics, specifically quantum field theory. This precision-LHC program rests on the comparison of measured scattering events with first-principle predictions provided by multi-purpose event generators, such as  \pythia~\cite{Sjostrand:2014zea}, \mg~\cite{Maltoni:2002qb,Alwall:2007st,Alwall:2011uj,Alwall:2014hca}, \sherpa~\cite{Sherpa:2019gpd}, and \herwig~\cite{Bewick:2023tfi}. Modern machine learning (ML)~\cite{Butter:2022rso,Plehn:2022ftl} allows us to benefit from the sizes of the observed and simulated datasets, so that we can extract all available information despite the complexity. Crucially, this acceleration has to cover the efficiency of simulation tools and of analysis workflows.

If we want to understand the LHC dataset in terms of quantum field theory, ML-enhanced event generators must be faster and more precise. This holds for new physics searches and for precision analyses, unified by the question: \textsl{Which Lagrangian describes the LHC data?} Specifically for \mg,  \madnis~\cite{Heimel:2022wyj,Heimel:2023ngj,Heimel:2024wph} sampling and uncertainty-aware amplitude surrogates~\cite{Bahl:2024gyt,Bahl:2025xvx,Bahl:2026qaf} provide a comprehensive ML basis to accelerate it to the point where, for instance, higher orders in perturbation theory and high-multiplicity final states are available at minimal additional computational cost. Similar ML improvements are on the way for \sherpa~\cite{janssen:2025zke,Herrmann:2025nnz}, \pythia~\cite{Butter:2025wxn}, and \herwig~\cite{Chan:2023icm}, forming a foundation of highly efficient computational tools for the HL-LHC program.

Given the availability of accelerated theory simulation tools, we now have to ensure that these tools can be used efficiently. Specifically, the next-generation computational structure of \mg7 has to be accompanied by ML-accelerated user training and support. This starts with the installation and includes learning-by-doing \mg runs for inexperienced users or incoming graduate students, as well as support for advanced users. The state of the art for such a support are physics chatbots, based on large language models (LLMs), often organized as agents. In this paper, we present \ma as a physics-specific agentic \mg support tool.

However, ML agents not only make established research or analysis tasks more convenient, they can accelerate them to the point where  ML agents transform the way we do research. It has already been illustrated how agents can perform, essentially, an entire experimental LHC analysis~\cite{Diefenbacher:2025zzn} or a theoretical particle physics calculation~\cite{Schwartz:2026ekw}. Following this path, we will show how \ma can run autonomous simulation campaigns, extracting the relevant information from given articles and from available web content. Such autonomous event generation will expand agentic analysis methods towards theory-based precision measurements and global LHC analyses.

This paper is not intended to provide new physics insights, but it changes the way we do LHC event generation. It combines the precision simulation capabilities of current and future \mg releases with the ease of the web-based \mg4~\cite{Alwall:2007st}.  In Sec.~\ref{sec:ma} we introduce the structure and implementation of \ma. In Sec.~\ref{sec:apps_install} we show how \ma install not only \mg, but also related codes such as \delphes and even \textsc{\root}. Next, in Sec.~\ref{sec:apps_train} we illustrate how \ma help inexperienced users by designing dedicated \mg tutorials, running simulations, and performing standard tasks such as event reweighting. In Sec.~\ref{sec:apps_adv} we turn to the support of experienced users, for instance, the simulation of top pair production beyond leading-order on-shell precision and understanding the underlying theory assumptions. Finally, in Sec.~\ref{sec:apps_auto} we illustrate how \ma run automatized simulation campaigns, reproducing a simulation from a publication without any further communication with the user. In the appendices we will give many more details, all of which are also available together with our public code base.
Apps.~\ref{app:improvements} and \ref{sec:claude} describe refinements developed after these demonstrations, including a Claude Code implementation with a self-improvement loop.

\section{Setup}
\label{sec:ma}

Large Language Models (LLMs)~\cite{zhao2023survey} are autoregressive, generative networks based on the transformer architecture. For language generation, a text is first encoded in a sequence of discrete tokens. The set of tokens is finite and typically has around $10^5$ elements. The transformer encodes a categorical distribution over the token space conditioned on a variable-length sequence of tokens. To generate text, we sample a token from this distribution and append it to an existing sequence. This is repeated until we sample a stop token. State-of-the-art LLMs are able to input multi-modal data~\cite{yin2024survey}, most commonly images. Those are first preprocessed with an additional network, typically a vision transformer, into image tokens which can be interleaved with other tokens.

LLMs can invoke tools~\cite{parisi2022talm, schick2023toolformer} as predefined functions, for which the LLM generates the arguments. This way, the LLM queries information and interacts with its environment. The latter is commonly implemented as a React loop~\cite{yao2022react}, where an LLM searches for a way to accomplish a user's goal, executes a tool, digests its output, and repeats this loop until the goal is fulfilled. Since tool calls and output are part of the LLM's context, it can correct errors~\cite{chen2023teaching} and teach itself to write and use software~\cite{wang2023voyager, cai2023large, yang2024swe}. LLMs perform better if they first output a reasoning sequence, the chain-of-thought, and only then provide the answer or call a tool~\cite{wei2022chain}. For some LLMs, the reasoning effort is configurable. 

LLMs can be instructed to specialize on specific tasks through a system prompt. It is prepended to the conversation and defines workflow, constraints, answering styles, etc. A set of task-specialized LLMs, called agents, have shown to outperform one general-purpose LLM~\cite{du2023improving, liu2023dynamic, wang2024survey}. To manage the agents, an orchestrating agent usually delegates task to specialized agents~\cite{hong2023metagpt, chen2024survey}. We will use this structure to support \mg users through \ma.

\subsection{LLMs for fundamental physics}

LLMs for physics research have a rich history dominated by astrophysical and cosmological applications. A first obvious LLM task is literature search. astroBERT~\cite{grezes2021buildingastrobertlanguagemodel} is the first LLM trained on 400,000 ADS papers, but without generative capabilities.  AstroLLaMA~\cite{Nguyen:2023nhp} is a generative LLaMA-2, fine-tuned on 300,000 cosmological arXiv abstracts and able to generate new abstracts. AstroLLaMa-Chat~\cite{UniverseTBD:2024pmh}, adds introductions and conclusions to the AstroLLaMA training dataset, allowing it to answer physics questions. The AstroMLab study~\cite{ting2024astromlab1winsastronomy} then compares a range of astronomical literature LLMs and introduces a benchmarking dataset with 4425 multiple-choice questions. An interesting question is how the commercial Claude LLMs compare to domain-specific solutions. Cosmosage~\cite{deHaan:2024ybs} starts from the public Mistral model and includes question-answer tuning. It is fine-tuned on astrophysical texts. In particle physics, FeynTune~\cite{Richmond:2025lzg} allows for arXiv searches including hep-ph, hep-th, and gr-qc abstracts. It uses standard  LoRa fine-tuning on Llama~3.1.  MARVEL~\cite{mukund2026marvel} provides evidence-based answers for queries related to the Laser Interferometer Gravitational-wave Observatory and builds a bridge from general literature research to proper expert support.

From a scientific and agent perspective, Mephisto~\cite{sun2025interpretingmultibandgalaxyobservations} is much more exciting, as it defines an agentic analysis of multi-band galaxy observations. It works with the latest James Webb data, in terms of spectral energy distributions, and includes Little Red Dot galaxies as a current research question. The LLM interacts with the CIGALE code base, develops the best parameter setting, and mimics scientific intuition about these parameters.  The CMBAgent~\cite{Laverick:2024fyh} works on Atacama Cosmology Telescope data with its public likelihood exists. It relies on the CLASS tool set to build a simulation and inference pipeline. 
It combines Retrieval Augmented Generation agents, coder agents, and manager agents, but relies on five successive steps with human feedback. Its result is a CMB inference for instance of $H_0$, $\sigma_8$, and $\Omega_c h^2$. The AI Cosmologist~\cite{Moss:2025ynt} combines planning, coding, execution, analysis, synthesis, and literature agents. Its output includes a research paper. SimAgents~\cite{Zhang:2025uoo} focuses on setting up simulations and processing the simulated data using a physics and a simulation agent. The simulations are checked using a visualization and a simple statistical analysis. It serves as a key inspiration for our \ma. Finally, L3M~\cite{Heneka:2025fpe} serves as a first example of how LLMs can be fine-tuned to process physics data with the help of simple connector networks.

The first particle physics agents are being developed right now. Experimental Agents for Discovery~\cite{Diefenbacher:2025zzn} automatize an anomaly search with the help of a researcher, a coder, a code reviewer, and a logic reviewer agent. The paper discusses in detail the prompting of an autonomous analysis agent. ArgoLOOM~\cite{Bakshi:2025fgx} combines particle physics, nuclear physics, and cosmology. On the particle physics side it runs \mg through an OpenAI API, where the UFO file allows it to go beyond particle physics. The Berkeley analysis agent~\cite{Gendreau-Distler:2025fsj} maximizes the expected significance for $H \to \gamma \gamma$ given simulated datasets. It focuses on standard experimental tasks, converting \root files and separating signal and background using a TabPFN transformer, and turning the outcome into a statistical statement. The paper compares different LLMs for each steps. Finally, HEPTAPOD~\cite{Menzo:2025cim} is a particle physics simulation agent that does not attempt to be autonomous. Given a model, it generates events using \mg and \pythia, reconstructs jets, and searches for resonances. 

For cosmological simulation tools it is becoming standard to provide LLM agents to help set up, run, and analyze the simulations, for instance, for CLASS/\href{https://github.com/santiagocasas/clapp/}{CLAP} and \href{https://github.com/franciscovillaescusa/CAMELS_Agents/}{CAMEL}. In particle physics, such simulation support agents do not exist yet.

\subsection{Structure}

\begin{figure}[b!]
  \centering
  \tikzset{
  mad box/.style={
    draw, thick, rounded corners=2pt,
    align=center,
    inner sep=5pt,
    outer sep=0pt,
    minimum height=7mm
  },
  mad user/.style={
    mad box, very thick, fill=red!12,
    font=\small,
    minimum height=10mm
  },
  mad orch/.style={
    mad box, thick, fill=red!12,
    font=\small,
    minimum width=33mm,
    minimum height=7mm
  },
  mad ctrl/.style={
    mad box, thick, fill=blue!7,
    font=\small,
    minimum width=33mm,
    minimum height=7mm
  },
  mad work/.style={
    mad box, thick, fill=green!7,
    font=\small,
    minimum width=33mm,
    minimum height=7mm
  },
  mad group/.style={
    draw, thick, rounded corners=4pt,
    inner sep=7pt,
    outer sep=0pt
  },
  mad title/.style={
    font=\small,
    fill=white,
    inner sep=1.5pt
  },
  mad flow/.style={
    <->, thick,
    line cap=round,
    line join=round,
    >=Latex
  }
}

\newdimen\MadAgentsWidth

\begin{tikzpicture}

\node[mad orch] (O) at (0,0) {Orchestrator};

\node[mad ctrl] (P)  at (-2.3, 1.55) {Planner};
\node[mad ctrl] (R)  at ( 2.3, 1.55) {Reviewer};
\node[mad ctrl] (PU) at ( 0.0,-1.75) {Plan-Updater};

\node[mad group, fit=(P)(R)(O)(PU)] (Gctrl) {};
\node[mad title, anchor=south west] at (Gctrl.north west) {Controlling agents};

\node[mad work] (W1) at ( 7.2,  2.25) {MG-Operator};
\node[mad work] (W2) at ( 7.2,  1.25) {Script-Operator};
\node[mad work] (W3) at ( 7.2,  0.25) {CLI-Operator};
\node[mad work] (W4) at ( 7.2, -0.75) {Plotter};
\node[mad work] (W5) at ( 7.2, -1.75) {PDF-Reader};
\node[mad work] (W6) at ( 7.2, -2.75) {Researcher};

\node[mad group, fit=(W1)(W6)] (Gwork) {};
\node[mad title, anchor=south west] at (Gwork.north west) {Worker agents};

\node[mad group, fit=(Gctrl)(Gwork), inner sep=14pt] (Gsys) {};

\path let \p1=(Gsys.west), \p2=(Gsys.east) in
  \pgfextra{%
    \global\MadAgentsWidth=\x2\relax
    \global\advance\MadAgentsWidth by -\x1\relax
  };

\node[mad user, minimum width=\MadAgentsWidth] (U) at ($(Gsys.north)+(0,1.05)$) {Human user};

\draw[mad flow] (O.north) -- (U.south -| O.north);

\draw[mad flow] (O.north west) -- (P.south);
\draw[mad flow] (O.north east) -- (R.south);
\draw[mad flow] (O.south) -- (PU.north);

\draw[mad flow] (O.east) -- (Gwork.west |- O);

\end{tikzpicture}
  \caption{\ma structure.}
  \label{fig:madagents-overview}
\end{figure}

The agentic structure of \ma is illustrated in Fig.~\ref{fig:madagents-overview}. It is built with \href{https://www.langchain.com/langgraph}{LangGraph} and uses the established orchestrating architecture. Instead of a strict sequence of agent actions, an Orchestrator decides the chain of agent calls. It chooses whether to message the user or to instruct a specific agent. For the latter, it decides which agent to call and generates an instruction. The selected agent carries out the instructed task and reports back to the Orchestrator. This allows \ma to adapt their response to the task and gives the user access to its workflow through instructions. We distinguish two main agent types:
\begin{enumerate}
\item \textbf{Controlling agents} manage and control the whole workflow, observe the user messages and the Orchestrator messages with the associated agent responses. The Orchestrator delegates straight-forward tasks directly to a Worker and otherwise instructs the Planner to generate a multi-step plan. For such a plan, the Orchestrator works on one plan step at a time and keeps the plan updated by invoking the Plan-Updater. Finally, the Orchestrator asks the Reviewer for its judgment after creating plans,  after worker agents produce user-facing results, after steps with expensive failure modes and before reporting final outcomes to the user. Depending on the Reviewer's response, the Orchestrator might decide to revise the step or the plan.
  \begin{itemize}
  \item Orchestrator fulfills the user request by invoking the specialized workers with suitable instructions; It interacts with the user, clarifies open or ambiguous user requests, delegates tasks to the specialized worker agents, and reports final results and answers. While the Orchestrator can in principle control the whole workflow, we support it with three specialized controlling agents;
  \item Planner generates a multi-step plan for solving a complex user request;
  \item Plan-Updater, a light-weight agent that only sees the Orchestrator instructions and translates them into updates of a given plan. 
  \item Reviewer verifies the worker results and the progress of the user request.
  \end{itemize}
\item \textbf{Worker agents} carry out requested tasks and only see a restricted context. Each Worker is conditioned only on their work, such that it sees the Orchestrator instructions, their internal workflows including tool calls, and their final replies to the Orchestrator.
  \begin{itemize}
  \item MG-Operator  specializes in the \mg software. We explicitly include an instruction for common use-patterns of \mg into its system prompt. It can modify and inspect the file system, run bash scripts, and access a dedicated interactive command line interface (CLI);
  \item Script-Operator specializes in bash and \python. Similarly to the MG-operator, its capabilities include modifying and inspecting the file system and running bash scripts. It should be used for CLI tasks that do not require using the \mg software;
  \item CLI-Operator accesses the user's interactive CLI. It can inspect the current state of the CLI and execute commands on user request;
  \item Plotter creates figures, following a default guidance, \eg use LaTeX for mathematical symbols, unless the user explicitly overwrites it in the prompt. Furthermore, it is instructed to iteratively inspect every figure and adjust if visibility is bad or plot requirements are not fulfilled;
  \item PDF-Reader extracts information from a pdf-file and can inspect references by searching the web;
  \item Researcher extracts information from the web.
  \end{itemize}
\end{enumerate}
The MG-Operator, the Script-Operator, the CLI-Operator, and the Plotter can search the web in case they do not know how to proceed or they cannot solve an error. For long context histories, we resort to the standard approach of summarizing the oldest message history with a Summarizer agent. We append this summary to the corresponding agent's system prompts. Details of this agent system are provided in App.~\ref{app:agentic_details}.

\subsection{Implementation}

Our agentic support system $(i)$ launches a containerized environment; $(ii)$ launches a local back-end and front-end server hosting \ma and enables web-browser communication; $(iii)$ launches an interactive command line session inside the container; and $(iv)$ attaches the user's command line terminal and \ma to this command line session. The user can work within the container without having to communicate with \ma. Once the user decides to do so, \ma can be invoked via the web-browser. The container, built and run via \href{https://apptainer.org/}{Apptainer}, hosts all applications, in particular \ma and \mg. We provide two pre-defined container setups:
\begin{enumerate}
\item A basic \mg installation containing \fastjet, \lhapdf, \pythia8, \delphes, and a basic set of \python packages. For standard \mg workflows, this image can be used for plug-and-play solutions.
\item Leaving out any \mg or \python related installations. This container can be used for version-specific installations which \ma can create.
\end{enumerate}
The directory structure includes:
\begin{itemize}
\item \lstinline|/output|: bind-mounted from the host file system to the container for the user work. \ma will save user-requested output there.
\item \lstinline|/runs/workdirs|: bind-mounted from the host file system to the container to store \ma internal data. A custom directory is assigned to each run, inside which the \ma workspace folder is located. The workspace, system-linked to \lstinline|/workspace|, is a directory dedicated to \ma for temporary and non-user-facing files.
\item \lstinline|/runs/pdf_files|: bind-mounted from the host file system to the container and system-linked to \lstinline|/pdf_files|, for users to place pdf files. Inside the container, this directory is read-only.
\item \lstinline|/opt|: to place software installations.
\end{itemize}
Any directory that is not bind-mounted to the host will still keep its contents between container runs. Apptainer stores those changes in the container’s overlay filesystem, so installs remain available across \ma invocations. 

\clearpage
\section{Agentic \mg}
\label{sec:apps}

There are several ways for \ma to improve and accelerate LHC research, from installing \mg and related software (including \root) in Sec.~\ref{sec:apps_install} to training-level support in Sec.~\ref{sec:apps_train}, experienced user support in Sec.~\ref{sec:apps_adv} and autonomous event generation in Sec.~\ref{sec:apps_auto}. Throughout our examples, we use the \ma implementation presented in Sec.~\ref{sec:ma} and \href{https://platform.openai.com/docs/models/gpt-5.1}{GPT-5.1} for all agents except the Plan-Updater, which uses \href{https://platform.openai.com/docs/models/gpt-5-mini}{GPT-5 mini}. We present individual MadAgents invocations without selecting for favorable outcomes and remind the reader that LLM responses are not deterministic. As \mg and the associated tools are open-source, basic instructions and basic knowledge are encoded in the trained LLM. This allows \ma to find and implement many plans and solutions without external information.

\subsection{Software Installation}
\label{sec:apps_install}

As a first step, we show how \ma can install \mg and associated public codes autonomously. We deliberately make this task hard by asking \ma to compile \root instead of downloading pre-compiled binaries. For this first example, we describe the \ma workflow in detail, including the plan creation and review steps. We start with the clean container environment and query,
\begin{userblock}[User]
Please install the latest stable release of MadGraph together with Pythia8 and Delphes under \lstinline|/opt|:
\begin{itemize}
    \item Install MadGraph under \lstinline|/opt/MG5_aMC| so that the main executable is located inside the directory \lstinline|/opt/MG5_aMC/bin|.
    \item Install Python dependencies into the MAD environment.
    \item Only install dependencies that are strictly required for a successful installation (i.e., missing them causes the install/build to fail). Install any such missing dependencies under \lstinline|/opt|. Don’t ask for additional details about dependency installation decisions; use your best judgment and proceed.\\
    If ROOT is required, build and install ROOT from source (compile it yourself) rather than using any precompiled/prebuilt ROOT binaries.
\end{itemize}

These installations must persist for future sessions.

\vspace{1em}
\hrule
\vspace{1em}

\textbf{Prompt explanation:} We tell \ma that this installation is not temporary and must persist for other \ma runs. This forces \ma to modify the persistent directories, for which \ma requires explicit permission. To skip this question, we specify the exact installation location. Furthermore, we give \ma direct permission to install dependencies, which otherwise might result in \ma asking follow-up questions.
\end{userblock}
For this complex task, \ma decides to create a plan which the reviewer checks and accepts,
\begin{planblock}[Plan]
\begin{enumerate}[leftmargin=*, label=\textbf{Step \arabic*:}, itemsep=0.6em]
\item Inspect existing environment and prerequisites
\item Download and install the latest stable MadGraph version
\item Install minimal required Python dependencies for MadGraph
\item Build and install ROOT from source
\item Install and integrate Pythia8 with MadGraph
\item Install and integrate Delphes with MadGraph
\item Finalize configuration, document paths, and run validation tests
\end{enumerate}
\end{planblock}
First, the Script-Operator inspects the container environment and \ma mark the first step as finished through the output
\begin{planblock}[\ma outcome of Step 1]
Environment under /opt inspected. Findings: /opt/MG5\_aMC does not exist; there is no existing MadGraph, Pythia, Delphes, or ROOT installation under /opt. /opt/envs/MAD exists and is a functional Python 3.11.2 environment with pip available. /opt/root does not exist. Core build tools gcc, g++, gfortran, make, cmake, git, tar, and gzip are all present in the system PATH. No filesystem modifications were made.
\end{planblock}
In the next two steps, \ma install \mg. The MG-Operator downloads and installs the \mg binaries. After the reviewer verifies the success, \ma report
\begin{planblock}[\ma outcome of Step 2]
MadGraph5\_aMC installed under /opt/MG5\_aMC. Version file reports version = 3.7.0, date = 2026-01-05. Standard MG5 tree present (aloha, madgraph, models, Template, bin, VERSION, etc.). /opt/MG5\_aMC/bin/mg5\_aMC exists, is a Python script with \#!/usr/bin/env python3 shebang, and has executable permissions. No extra system or Python packages were installed in this step. Basic invocation of mg5\_aMC using the /opt/envs/MAD Python 3.11.2 environment fails only due to missing Python module six; no other installation or permission errors observed. Persistent files confined to /opt/MG5\_aMC; temporary artifacts under /workspace/mg5\_src.
\end{planblock}
The Script-Operator installs the missing dependencies. Once \mg can start, the reviewer judges the outcome and \ma note
\begin{planblock}[\ma outcome of Step 3]
Configured /opt/envs/MAD (Python 3.11.2, pip 25.3) so that MadGraph5\_aMC runs without Python import errors. Using /opt/envs/MAD/bin/python to run /opt/MG5\_aMC/bin/mg5\_aMC initially failed due to missing module six. Installed only the required package six via '/opt/envs/MAD/bin/python -m pip install six' (six==1.17.0 in /opt/envs/MAD/lib/python3.11/site-packages). After installation, MG5 3.7.0 starts under /opt/envs/MAD and executes simple command files (e.g. set automatic\_html\_opening False; quit) without any ImportError or ModuleNotFoundError. Remaining messages are only warnings about external tools (fastjet-config, eMELA-config, lhapdf-config, EPS viewer, browser), not Python dependency issues. No other Python packages were added in this step.
\end{planblock}
Now, \ma target the \root installation as the most challenging step. Mentioning \root in the prompt reminds \ma that \root is required by \delphes. If we omit the explicit \root instruction, \ma will most likely not have a planned step to install \root, encounter an error while installing \delphes, trace the error to a missing \root installation, and install it then. 

\ma start by inspecting the operating system and the installed compilers. They download  \root from the official source to \lstinline|/workspace/root_build|, unpack it, create a build directory, configure a minimal \root build for \delphes, and compile it with parallel jobs. After a few minutes, the bash execution is sent to the background and its \lstinline|stderr| and \lstinline|stdout| streams are forwarded to files. This allows \ma to call tools in parallel. \ma track the compilation by periodically checking the bash output and putting themselves to sleep for a few minutes. After successful compilation, they install \root on \lstinline|/opt/root| and verify the installation by inspecting the \root directory and executing \lstinline|root-config|. This installation is judged by the reviewer, and \ma summarize\footnote{Throughout the paper we use [\dots] to indicate a shortened \ma prompt or output} 
\begin{planblock}[\ma outcome of Step 4 (shortened)]
Built and installed ROOT 6.30.06 from the official source tarball (\url{https://root.cern/download/root_v6.30.06.source.tar.gz}) using CMake and make. [\ldots]\ thisroot.sh and related env scripts are present in /opt/root/bin. No prebuilt ROOT binaries or external ROOT packages were used; ROOT was built entirely from source and installed under /opt/root without adding extra external dependencies beyond ROOT’s builtin third-party components.
\end{planblock}
Next, the MG-Operator installs \pythia and \delphes, each using the \mg installation function. Both installations are verified by the reviewer, and \ma note
\begin{planblock}[\ma outcome of Step 5 (shortened)]
Installed and integrated Pythia8 with MadGraph5\_aMC under /opt using MG5’s internal HEPTools installer. Ran MG5 in batch mode via /opt/envs/MAD to execute 'install pythia8', which downloaded and built Pythia8 and related HEPTools under |/opt/MG5\_aMC/HEPTools. [\ldots] \ 
MG5aMC\_PY8\_interface was installed under /opt/MG5\_aMC/HEPTools/MG5aMC\_PY8\_interface. [\ldots] A test MG5 run [\ldots] executed via /opt/envs/MAD completed successfully: [\ldots]
\end{planblock}
and
\begin{planblock}[\ma outcome of Step 6 (shortened)] 
Installed and integrated Delphes with MadGraph5\_aMC using ROOT from /opt/root. Initial MG5 'install Delphes' failed due to MG5 requiring a 'root' executable on PATH; our ROOT build under /opt/root/bin provided 'root.exe' and 'root-config' but no 'root'. Created a minimal symlink /opt/root/bin/root -> root.exe to satisfy MG5's check. [\ldots] re-ran 'install Delphes' via MG5's installer under /opt/MG5\_aMC. Delphes was built successfully and installed at /opt/MG5\_aMC/Delphes [\ldots]  A functional MG5 test chain [\ldots] run via /opt/envs/MAD with the ROOT environment set as above, completed successfully: [\ldots]
\end{planblock}
\ma now finalize the installation with the Script-Operator. As this step is straight-forward, it does not require an additional reviewer invocation and \ma marks the final step as accomplished with the outcome
\begin{planblock}[\ma outcome of Step 7 (shortened)]
Final validation and summarization completed. Confirmed that /opt/MG5\_aMC/bin/mg5\_aMC exists, is executable, and runs cleanly under the /opt/envs/MAD\ Python environment for a trivial command file [\ldots], with only expected external-tool warnings and no Python import errors. Collected version information: MadGraph5\_aMC version 3.7.0 (from /opt/MG5\_aMC/VERSION, date 2026-01-05); ROOT version 6.30.06 (from \/opt/root/bin/root-config --version); Pythia8 version 8.316 (from /opt/MG5\_aMC/HEPTools/pythia8/bin/pythia8-config --version); Delphes version 3.5.0 (from \/opt/MG5\_aMC/Delphes/CHANGELOG). Verified key installation paths: [\ldots] Confirmed these directories exist and have expected top-level contents. [\ldots]
\end{planblock}
Before reporting back to the user, \ma ask the reviewer to judge whether the user goal has been achieved completely. After the reviewer confirms this, \ma reports
\begin{finalblock}[\ma (shortened)]
Here is what is now installed and how to use it.

\textbf{1. MadGraph5\_aMC}

\begin{itemize}
\item Install location: \lstinline|/opt/MG5_aMC|
\item Version: 3.7.0 [...]
\item Main executable: \lstinline|/opt/MG5_aMC/bin/mg5_aMC| [...]
\item {[...]}
\end{itemize}

\textbf{How to run MG5}

Use the MAD environment’s Python explicitly, e.g.

\begin{bashblock}
cd /opt/MG5_aMC
PYTHONUNBUFFERED=1 /opt/envs/MAD/bin/python /opt/MG5_aMC/bin/mg5_aMC
\end{bashblock}
[...]

\textbf{2. Python environment (\lstinline|/opt/envs/MAD|)}

\begin{itemize}
\item Path: \lstinline|/opt/envs/MAD|
\item Python: 3.11.2
\item {[...]}
\item Only MG5-specific addition: \lstinline|six|
\end{itemize}

\textbf{3. ROOT (for Delphes)}

\begin{itemize}
\item Built \textbf{from source} (no prebuilt binaries) from: \url{https://root.cern/download/root_v6.30.06.source.tar.gz}
\item Install prefix: \lstinline|/opt/root|
\item Version: 6.30.06
\item {[...]}
\end{itemize}

\textbf{Typical ROOT environment setup (if you need ROOT directly)}

[...]
\begin{bashblock}
source /opt/root/bin/thisroot.sh
\end{bashblock}
[...]

\textbf{4. Pythia8}

\begin{itemize}
\item Installed using MG5’s \lstinline|install pythia8| [...]
\item Version: 8.316 [...]
\item Location: \lstinline|/opt/MG5_aMC/HEPTools/pythia8| [...]
\item {[...]}
\item Additional HEPTools components built and kept under \lstinline|/opt/MG5_aMC/HEPTools/| (by MG5’s installer): HEPMC, LHAPDF (Python 3), zlib, etc.
\end{itemize}
[...]

\textbf{5. Delphes}

\begin{itemize}
\item Installed via MG5’s \lstinline|install Delphes| [...]
\item Location: \lstinline|/opt/MG5_aMC/Delphes| [...]
\item Version: 3.5.0. [...]
\item {[...]}
\end{itemize}

\textbf{Test performed (MG5 + Pythia8 + Delphes)}

MG5 command file (schematically):

\begin{bashblock}
set automatic_html_opening False
import model sm
generate e+ e- > mu+ mu-
output /workspace/mg5_test_delphes_output
launch /workspace/mg5_test_delphes_output
shower=PYTHIA8
detector=Delphes
0
quit
\end{bashblock}
[...]

Result: parton-level generation, Pythia8 showers, and the Delphes step all completed without ROOT/Delphes errors (10k events, consistent cross section). [...]
\end{finalblock}

\clearpage
\subsection{Supporting and training inexperienced users}
\label{sec:apps_train}

Our second \ma application is not an autonomous task, but training and support of inexperienced users. Here, it is crucial that the \ma communicate what they are doing, where problems appear, and how they solve these problems. We provide three levels of support, from tailored tutorials in Sec.~\ref{sec:apps_train_tut} to learning-by-doing in Sec.~\ref{sec:apps_train_learn} and comprehensive documentation of more involved tasks in Sec.~\ref{sec:apps_train_doc}.

\subsubsection{Tailored tutorials}
\label{sec:apps_train_tut}

We begin by asking \ma to train an inexperienced user interactively and through tutorials tailored to their research tasks,

\begin{userblock}[User]
I am a physics PhD student and completely new to MadGraph. My task is to generate LHC events for BSM models (expressed as an EFT) and then compare the generated events to real data via SBI.

Be my MadGraph teacher. Please create tutorials in \lstinline|/output/tutorials| where I learn the basics of generating the relevant data via MadGraph (my supervisor told me that Pythia and Delphes suffice for the beginning). I should solve the tutorials in \lstinline|/output/tutorials/my_work/<dedicated_tutorial_folder>|. I will inspect the tutorials with VS Code and work through them via my interactive CLI that you have access to as well.

When creating the tutorials: Do not save or state the solutions to the tutorials anywhere, not even inside the worker's execution trace.

In your reply, include an overview of the tutorials you created: for each tutorial, give a short description and explain what I should learn from it.

\vspace{1em}
\hrule
\vspace{1em}

\textbf{Prompt explanation:} We inform \ma of our situation, our current knowledge and the future task that it must prepare us for. We define \ma's teaching role, specify which directories it should modify and explicitly mention that we will work through the tutorials in the container environment. This allows \ma to not only tailor the tutorials on our research task, but also on the environment. Finally, we instruct \ma to not disclose any solution at this stage and to include an overview of the created tutorials.
\end{userblock}
\ma develop a plan and work through it to create the tutorials
\begin{planblock}[Plan]
\begin{enumerate}[leftmargin=*, label=\textbf{Step \arabic*:}, itemsep=0.6em]
\item \textbf{Design tutorial sequence and structure:}\\
\textbf{Outcome:} Designed a sequence of 6 tutorials with folder names, titles, prerequisites, learning objectives, and open-ended exercise prompts; saved detailed design in /workspace/madgraph\_tutorials/tutorial\_design\_overview.md.\\
\textbf{Agents:} Script-Operator

\item \textbf{Create tutorial tree for tutorials and user work:}\\ 
\textbf{Outcome:} Created /output/tutorials with six tutorial subdirectories and a top-level index README, plus mirrored work directories under /output/tutorials/my\_work with minimal README placeholders; no solutions or numeric answers were added.\\
\textbf{Agents:} Script-Operator

\item \textbf{Write introductory MG5/Pythia/Delphes tutorials (basics and SM processes):}\\ 
\textbf{Outcome:} Created detailed beginner-friendly README tutorials in /output/tutorials/01\_mg5\_basics and /output/tutorials/02\_sm\_drell\_yan\_parton covering MG5 basics, the MG5→Pythia→Delphes workflow, and SM Drell–Yan at parton level with open-ended exercises and no solutions or numeric answers.\\
\textbf{Agents:} MG-Operator $\rightarrow$ Reviewer

\item \textbf{Write tutorials on Pythia showering and Delphes detector simulation:}\\ 
\textbf{Outcome:} Created detailed tutorials in /output/tutorials/03\_pythia\_showering and /output/tutorials/04\_delphes\_detector covering Pythia showering and Delphes detector simulation, with open-ended exercises and no solutions or numeric answers.\\
\textbf{Agents:} MG-Operator $\rightarrow$ Reviewer

\item \textbf{Write tutorials on BSM EFT models and SBI-oriented outputs:}\\ 
\textbf{Outcome:} Created detailed tutorials in /output/tutorials/05\_bsm\_eft\_models and /output/tutorials/06\_sbi\_dataset\_preparation covering EFT UFO models in MG5, EFT parameter scans, and SBI-oriented dataset and metadata design, with open-ended exercises and no solutions or numeric answers.\\
\textbf{Agents:} MG-Operator $\rightarrow$ Reviewer

\item \textbf{Consistency and quality review, then provide user-facing overview:}\\ 
\textbf{Outcome:} Performed a light automated consistency scan over all six tutorial README files to check for structural issues (forbidden wording, correct my\_work references, path sanity). No problematic numeric answers or explicit solutions were detected under these checks. Created a high-level, solution-free overview of all tutorials at /workspace/madgraph\_tutorials/tutorials\_overview\_for\_user.md, summarizing titles, content, and learning outcomes.\\
\textbf{Agents:} Script-Operator
\end{enumerate}
\end{planblock}
After verifying that the user goal is completed, \ma replies
\begin{promptblock}[\ma (shortened)]
I’ve created a 6-part tutorial suite under \lstinline|/output/tutorials|, with matching work areas under \lstinline|/output/tutorials/my_work|. Below is an overview of each tutorial: what it’s called, where it lives, and what you’ll learn. [...]

\vspace{1em}
\hrule
\vspace{1em}

\textbf{Tutorial 1 – MG5 basics and the full pipeline}
\begin{itemize}
 \item \textbf{What it covers:}
 \begin{itemize}
 \item What MadGraph5\_aMC@NLO is and how it fits into the LHC simulation chain.
 \item The overall workflow: MG5 (matrix elements, parton level) $\to$ Pythia (showers, hadronization) $\to$ Delphes (detector level).
 \item Installing/obtaining MG5\_aMC (at a generic, OS-agnostic level) and launching the \lstinline|mg5_aMC| interface.
 \item Basic MG5 commands (\lstinline|import model|, \lstinline|generate|, \lstinline|output|, \lstinline|launch|, \lstinline|display ...|) and the structure of a process directory (\lstinline|Cards/|, \lstinline|Events/|, etc.).
  \end{itemize}
\item {[...]}
\end{itemize}

\vspace{1em}
\hrule
\vspace{1em}

\textbf{Tutorial 2 – SM Drell–Yan at parton level}
\begin{itemize}
 \item \textbf{What it covers:}
\begin{itemize}
\item The physics of Drell–Yan at the LHC (e.g. processes like $pp \to \ell^+\ell^-$) as a benchmark.
\item Defining Drell–Yan–like processes in MG5 using the Standard Model and multiparticle labels for protons and leptons.
\item Using \lstinline|run_card.dat| (beams, PDFs, cuts, number of events, etc.) and \lstinline|param_card.dat| (SM parameters) in a concrete example.
\item Running a parton-level Drell–Yan sample, locating LHE files, and qualitatively inspecting events.
\end{itemize}
\item {[...]}

\end{itemize}

\vspace{1em}
\hrule
\vspace{1em}

\textbf{Tutorial 3 – Pythia: from partons to hadrons}
\begin{itemize}
\item \textbf{What it covers:}
\begin{itemize}
\item The roles of initial-state radiation, final-state radiation, hadronization, and underlying event in Pythia.
\item How to start from an MG5 process directory with LHE events (e.g. Drell–Yan) and enable Pythia showering via the MG5 run interface.
\item Where Pythia configuration lives (shower card, tunable settings, seeds) in a typical MG5 setup.
\item Where hadron-level outputs are written (alongside LHE files) and how they differ conceptually from parton-level events.
\end{itemize}
\item {[...]}
\end{itemize}

\vspace{1em}
\hrule
\vspace{1em}

\textbf{Tutorial 4 – Delphes: fast detector simulation}
\begin{itemize}
\item \textbf{What it covers:}
\begin{itemize}
\item The purpose of fast detector simulation and where Delphes sits in the pipeline.
\item Delphes detector cards: what they control (acceptance, resolutions, object definitions, etc.).
\item Running Delphes via MG5’s run interface on Pythia-showered events, and the basic idea of standalone Delphes usage.
\item The structure of Delphes ROOT outputs (collections for reconstructed electrons, muons, jets, MET, etc.) and how to start inspecting them.
\end{itemize}
\item {[...]}
\end{itemize}

\vspace{1em}
\hrule
\vspace{1em}

\textbf{Tutorial 5 – BSM EFT models in MadGraph}
\begin{itemize}
\item \textbf{What it covers:}
\begin{itemize}
\item Conceptual introduction to BSM EFTs [...]
\item UFO models in MG5: how BSM/EFT models encode new particles, interactions, and parameters.
\item Importing an EFT UFO model, inspecting new particles/parameters/interactions, and locating EFT parameters in \lstinline|param_card.dat|.
\item Defining simple EFT-sensitive processes [...] and setting up runs.
\item Organizing parameter scans over EFT coefficients and recording metadata [...]
\end{itemize}

\item \textbf{What you should learn:}
\begin{itemize}
\item How to work with EFT/BSM UFO models in MG5 instead of the default SM.
\item How to identify and modify EFT parameters/Wilson coefficients in \lstinline|param_card.dat|.
\item How to set up and run simple EFT-deformed processes at the LHC.
\item How to design and document basic EFT parameter scans in a way that is reproducible and ready for later analysis/SBI.
\end{itemize}
\item {[...]}
\end{itemize}

\vspace{1em}
\hrule
\vspace{1em}

\textbf{Tutorial 6 – Preparing datasets for SBI}
\begin{itemize}
\item \textbf{What it covers:}
\begin{itemize}
\item A conceptual view of Simulation-Based Inference (SBI): [...]
\item Designing a directory structure for simulations over parameter points: [...]
\item Designing metadata [...]
\item Conceptual discussion of event-level vs.\ summary-statistic representations and how they might be used for SBI (without implementing any SBI algorithms).
\item Strategies for train/validation/test splits in parameter space and for ensuring reproducibility.
\end{itemize}
\item {[...]}

\end{itemize}

\vspace{1em}
\hrule
\vspace{1em}

\textbf{Where to look and how to proceed}
\begin{itemize}
\item All tutorial instructions are in: [...]
\item Your work areas (where you should edit files, run commands, and keep results) are: [...]
\end{itemize}

A good way to get started is to open \lstinline|/output/tutorials| in VS Code, read through \lstinline|01_mg5_basics/README.md|, and then begin working in \lstinline|my_work/01_mg5_basics| from your CLI. When you’re comfortable with MG5 basics and parton-level Drell–Yan, continue in order through Pythia, Delphes, EFT, and finally the SBI dataset design.

If you’d like, I can next help you plan how to tackle these tutorials over time (e.g. which ones to do before you start touching real data or SBI code).
\end{promptblock}
\ma accompany the user through the \href{https://github.com/MadGraphTeam/MadAgents/tree/main/supplementary/tutorials}{tutorials}, provide detailed explanations and hints, and verify solutions. As an example, we work through the 5th tutorial, shown in App~\ref{app:ex_tutorial}. The first exercise requires loading a UFO model and inspecting it in \mg. After looking at the available UFO models in \mg, we ask \ma for help choosing a model
\begin{userblock}[User]
I skip tutorials 1–4 and go directly to tutorial 5. I see many UFO models and I don’t know which one to load for this tutorial. Please help me.
\end{userblock}
\ma reads the output of the user's CLI and responds
\begin{promptblock}[\ma (shortened)]
There is no single ``correct'' UFO for Tutorial 5 -- the tutorial is about \textit{how} to work with an EFT UFO, not about a specific physics choice. But given the models you have available, here is a pragmatic way to proceed.

\textbf{1. Quick sanity check with your supervisor}

If your project already has a specific EFT in mind (e.g.\ a particular top EFT, VBS EFT, DM EFT, etc.), the safest is to ask your supervisor which UFO name to use. That will align your tutorial work with your actual analysis.

\textbf{2. If you just want a generic SMEFT for learning}

From your \lstinline|display modellist| output you have several SMEFT/EFT UFOs, for example:
\begin{itemize}
\item \lstinline|SMEFTsim_A_general_MwScheme_UFO_v2|
\item \lstinline|SMEFTsim_general_MwScheme_UFO|
\item {[...]}
\end{itemize}

For a \textbf{generic, reasonably standard SMEFT playground}, I suggest you start with:

\begin{bashblock}
import model SMEFTsim_A_general_MwScheme_UFO_v2
\end{bashblock}

Reason:
\begin{itemize}
\item It’s a widely used SMEFT implementation.
\item It includes a broad set of dimension-6 operators, so you can practice importing the model, inspecting new parameters, and changing Wilson coefficients.
\item It’s not tied to one very narrow corner of parameter space (like 4-top only).
\end{itemize}

[...]

\textbf{3. How this ties into Tutorial 5}

Once you pick one of these models: [...]

If you tell me which physics sector your project cares about (top/EW/VBS/DM/\dots), I can narrow this further and point you to a more targeted model from your list.
\end{promptblock}
After loading the model and defining the Drell-Yan process, we ask \ma to help us understand the BSM-operators
\begin{userblock}[User]
I have loaded the model SMEFTsim\_A\_general\_MwScheme\_UFO\_v2 and generated a Drell–Yan-like process. However, when I inspect \lstinline|param_card.dat|, I’m overwhelmed by the sheer number of parameters. Please find an interesting BSM parameter (for this tutorial) and explain the physics behind it.
\end{userblock}
The \ma provide us with a guide to understand the UFO parameter cards,
\begin{promptblock}[\ma (shortened)]
A good ``first'' BSM parameter for a Drell--Yan tutorial in \lstinline|SMEFTsim_A_general_MwScheme_UFO_v2| is one of the \textbf{four-fermion contact operators that couple quarks and leptons}, e.g.\ the coefficient of the operator usually called
\[
\mathcal O_{lq}^{(3)} \sim (\bar l_L \gamma_\mu \tau^I l_L)(\bar q_L \gamma^\mu \tau^I q_L),
\]
where $l_L$ is the lepton doublet and $q_L$ is the quark doublet.

\textbf{1. Why this is a good choice for Drell--Yan}

For a process like
\[
pp \to \ell^+ \ell^-
\]
in the SM you mainly have $\gamma^*/Z$ exchange:
\begin{itemize}
\item $q \bar q \to \gamma^*/Z \to \ell^+ \ell^-.$
\end{itemize}

When you turn on an $\mathcal O_{lq}^{(3)}$--type coefficient:
\begin{itemize}
\item You add a \textbf{contact interaction} of the form
\[
(\bar q_L \gamma_\mu q_L)(\bar \ell_L \gamma^\mu \ell_L)
\]
(up to isospin structure),
\item This contributes a new \textbf{tree-level amplitude} that interferes with the SM $\gamma^*/Z$ exchange.
\end{itemize}

[...]

\textbf{2. How to locate such a parameter in the \lstinline|param_card.dat|}

In \lstinline|SMEFTsim_A_general_MwScheme_UFO_v2| the EFT coefficients are organized by \textbf{operator type and generation indices}. Exact names depend on the implementation, but:
\begin{itemize}
\item Look for a \textbf{BLOCK} that collects SMEFT coefficients (often something like \lstinline|BLOCK SMEFT| or blocks with names containing \lstinline|CLQ3|, \lstinline|CLQ1|, etc.).
\item Inside that block you’ll find entries corresponding to $O_{lq}^{(1)}$ and $O_{lq}^{(3)}$ with generation indices. For Drell--Yan to light charged leptons, you care about:
\begin{itemize}
\item quark generations 1 and 2 (u, d, s, c) in the proton,
\item lepton generations 1 and 2 (e, $\mu$).
\end{itemize}
\end{itemize}

[...]

\textbf{3. What to do with it in the tutorial (qualitatively)}

Once you’ve identified such a coefficient: [...]
\end{promptblock}
In case the user cannot locate such an parameter, \ma can be asked to help:
\begin{userblock}[User]
I am unable to find those operators in the UFO model.... Please find them for me.
\end{userblock}
\ma invoke the Script-Operator to locate the given parameter and responds
\begin{promptblock}[\ma (shortened)]
A concrete, interesting choice for Drell--Yan in \lstinline|SMEFTsim_A_general_MwScheme_UFO_v2| is one of the \textbf{four-fermion lepton--quark contact operators}. A representative example (as it appears in the SMEFTsim implementation) is a coefficient called
\begin{itemize}
\item \textbf{\lstinline|clq1Ph1112|} in \textbf{\lstinline|BLOCK FRBLOCK|} of the \lstinline|param_card.dat|.
\end{itemize}

\textbf{1. How to find it in your \lstinline|param_card.dat|}

[...]

\textbf{2. What \lstinline|clq1Ph1112| means physically}

\lstinline|clq1Ph...| coefficients belong to the operator family usually called
\[
O_{lq}^{(1)} \sim (\bar l_L \gamma_\mu l_L)(\bar q_L \gamma^\mu q_L),
\]
where $l_L$ is the left-handed lepton doublet and $q_L$ is the left-handed quark doublet. The suffix \lstinline|1112| encodes a particular light-generation flavor combination (first-generation leptons and light quarks) in the SMEFTsim naming scheme.

For neutral-current Drell--Yan $pp \to \ell^+\ell^-$:
\begin{itemize}
\item In the SM the dominant amplitude is $q\bar q \to \gamma^*/Z \to \ell^+\ell^-.$
\item Turning on \lstinline|clq1Ph1112| adds a \textbf{contact interaction} $(\bar q_L \gamma_\mu q_L)(\bar \ell_L \gamma^\mu \ell_L)$.
\item This new amplitude \textbf{interferes} with the SM $\gamma/Z$ exchange.
\end{itemize}

Qualitatively, this has two main effects:
\begin{enumerate}
\item \textbf{High-mass tail enhancement/distortion:} [...]
\item \textbf{Angular-distribution changes:} [...]
\end{enumerate}

\end{promptblock}
After launching three \mg runs with different \lstinline|clq1Ph1112| values, we ask \ma to verify the correctness of the users steps
\begin{userblock}[User]
I have executed three runs in which I changed the value of `clq1Ph1112`. Please verify that I did this correctly.    
\end{userblock}
\ma inspect the user's CLI and the run-specific directories and confirm
\begin{promptblock}[\ma]
You did it correctly.

\begin{itemize}
\item Your Drell--Yan process directory is \lstinline|DY_EFT| under \lstinline|/output/tutorials/my_work/05_bsm_eft_models| and it indeed uses the model \lstinline|SMEFTsim_A_general_MwScheme_UFO_v2|.
\item The master \lstinline|Cards/param_card.dat| in \lstinline|DY_EFT| contains \lstinline|clq1Ph1112| (labelled \lstinline|clq1ph1112| in \lstinline|BLOCK FRBLOCK|), so you edited the right parameter.
\item For the three runs \lstinline|run_01|, \lstinline|run_02|, and \lstinline|run_03| under \lstinline|DY_EFT/Events/|:
\begin{itemize}
\item Each run’s configuration (stored in its \lstinline|run_0x_tag_1_banner.txt|) has an entry for \lstinline|clq1ph1112|.
\item The value of \lstinline|clq1Ph1112| is different in each of the three runs.
\end{itemize}
\end{itemize}

So you have three distinct runs of the same Drell--Yan EFT process that differ only in the chosen value of \lstinline|clq1Ph1112|, exactly as intended for the tutorial. If you like, next we can discuss how to set up a simple comparison of the dilepton invariant-mass distributions between these runs (staying at a qualitative level).
\end{promptblock}

\subsubsection{Learning by doing}
\label{sec:apps_train_learn}

In case a user knows basic \mg, the \ma can lead them through their task and provide relevant information while working on it. For example, we want to generate events for the process
\begin{align}
    p p \to W^+ H 
    \to (\ell^+ \nu_\ell) \, (b \bar{b}) 
    \qquad \text{with} \qquad 
    \ell^+ = e^+, \mu^+ \; .
\end{align}
For a first \mg attempt, we execute
\begin{bashblock}
define l+ = e+ mu+
define vl = ve vm
generate p p > w+ h, (w+ > l+ vl) (h > b b~)
\end{bashblock}
In the last line, a comma is missing between both decay processes, leading to a \mg error. We query \ma

\begin{userblock}[User]
Why do I get this error?
\end{userblock}
\ma checks if this vague statement might refer to the CLI. They find and analyze the error and give the solution

\begin{promptblock}[\ma]
You get that error because of a syntax problem in the \texttt{generate} line, not because of the physics of the process.

You wrote:
\begin{bashblock}
generate p p > w+ h, (w+ > l+ vl) (h > b b~)
\end{bashblock}

After the main process \texttt{p p > w+ h}, \mg expects \textbf{each decay block to be separated by commas} and to be a valid process on its own. Here there is a missing comma between the two decay blocks, so MG5 tries to parse them as one malformed block and hits the deprecated \texttt{> A A >} pattern, giving the
\begin{bashblock}
InvalidCmd : Invalid "> A A >" syntax
\end{bashblock}
message.

Use one of these equivalent, valid syntaxes:
\begin{bashblock}
generate p p > w+ h, w+ > l+ vl, h > b b~
\end{bashblock}
or
\begin{bashblock}
generate p p > w+ h, (w+ > l+ vl), (h > b b~)
\end{bashblock}

Both correspond to $pp \to W^+ H$ with $W^+ \to \ell^+ \nu_\ell$ and $H \to b\bar b$.

\end{promptblock}

As a second example, we want to generate events for the process 
\begin{align}
    p p \to Z H 
    \to (e^+ e^-) \, (Z Z) \; .
\label{eq:def_hz}
\end{align}
In the Standard Model, this Higgs decay is not possible on-shell, but in the \mg workflow the Feynman diagram and process generation are independent of the value used for the Higgs mass. A more in-depth \ma discussion of off-shell decays will be part of Sec.~\ref{sec:apps_adv}. For now, we use \mg to execute
\begin{bashblock}
generate p p > z h, (h > z z, z > e+ e-)
\end{bashblock}
Using this command, the produced $Z$ does not decay at all, but the Higgs decays to $Z Z \to 4 e$. We first ask \ma to check whether the generated process matches the process in Eq.\eqref{eq:def_hz}:
\begin{userblock}[User]
Please verify whether my defined process corresponds to the process $p p \to Z H$, where the produced $Z$ decays to $Z \to e^+ e^-$ and the produced H to $H \rightarrow Z Z$.
\end{userblock}
\ma inspect the user's CLI, decide that the user request refers to it, and explains the difference,
\begin{promptblock}[\ma]
From your MG5 transcript, the process you defined is

\begin{bashblock}
generate p p > z h, (h > z z, z > e+ e-)
\end{bashblock}

This corresponds to

\begin{itemize}[label=\textbullet]
  \item production: \(pp \to Z_\text{assoc} H\)
  \item Higgs decay: \(H \to Z Z\)
  \item \textbf{one of the Z bosons from the Higgs decay} then decays as \(Z \to e^+ e^-\),
  \item while the \textbf{associated Z from production remains undecayed}.
\end{itemize}

So it does \textbf{not} exactly match the verbal description “\(pp \to ZH\) with the produced \(Z \to e^+ e^-\) and the produced \(H \to ZZ\) (with those Z bosons left stable)”. Instead, it decays a Z coming from the Higgs.

To get what you described (associated Z leptonically decaying, Higgs to ZZ with both Z from H stable), you should use

\begin{bashblock}
generate p p > z h, (z > e+ e-), (h > z z)
\end{bashblock}

Here:
\begin{itemize}[label=\textbullet]
  \item \texttt{(z > e+ e-)} attaches the decay to the \textbf{associated Z} in \texttt{p p > z h}.
  \item \texttt{(h > z z)} gives \(H \to Z Z\) with both Z’s left undecayed.
\end{itemize}

If you later also want to decay the Higgs Z’s, you’d extend the Higgs decay chain, e.g.

\begin{bashblock}
generate p p > z h, (z > e+ e-), (h > z z, z > e+ e-, z > mu+ mu-)
\end{bashblock}

which corresponds to \(Z_\text{assoc} \to e^+ e^-\) and the two Z’s from H going to \(e^+ e^-\) and \(\mu^+ \mu^-\), respectively.
\end{promptblock}
In the last step \ma go beyond the requested process and propose a natural refinement of our initial simulation. 

However, in the third bullet point they wrongly claim that only one $Z$ boson decays leptonically, where in truth both do. We discuss this mistake in App.~\ref{app:madgaph_mistakes} and offer a solution in which \ma is only allowed to make \mg statements if supported by evidence.

\subsubsection{Documented reweighted simulation}
\label{sec:apps_train_doc}

Yet another way \ma support users is a human-readable documentation of complex tasks. We use an example from the SFitter workflow~\cite{Elmer:2023wtr}:
\begin{enumerate}
    \item generate two parton-level $t \bar{t}$ datasets, one at LO and one at NLO;
    \item For each dataset, bin the events in $m_{t \bar{t}}$ and compute a bin-wise $K$-factor by reweighting events from LO to NLO;
    \item generate LO events including a chromomagnetic dimension-6 operator, including showering and detector effects;
    \item reweight those events with the $K$-factor determined for the SM-process.
\end{enumerate}
A single bash script controls the generation chain. The corresponding scripts are available in the \href{https://github.com/MadGraphTeam/MadAgents/tree/main/supplementary/dataset_documentation}{code repository}. We invoke this script and save the bash output to a log file. Apart from this, the dataset directory does not contain any documentation and the scripts do not contain comments. We query \ma to create such a  documentation

\begin{userblock}[User]
I have generated the dataset \lstinline|/output/dataset/output/dataset/ufo_ttbar_reweighted.npz| by running \lstinline|/output/dataset/run.sh|.

\textbf{Task:} Create a technically precise documentation in \lstinline|/output/dataset/documentation| by inspecting the directory \lstinline|/output/dataset|.

\textbf{Deliverables:}
\begin{enumerate}
\item \lstinline|/output/dataset/documentation/documentation.md| (detailed)
\item \lstinline|/output/dataset/documentation/summary.md| (short summary of the detailed documentation)
\end{enumerate}

The detailed documentation \textbf{MUST} include:

\textbf{A. Reproducibility / software environment}
\begin{itemize}
\item Core runtime(s) used (e.g., Python/R/Julia/MATLAB) + versions
\item Libraries/frameworks/external tools/dependencies used by dataset generation + versions
\end{itemize}

\textbf{B. Physics behind the dataset}
\begin{itemize}
\item What collider process(es) are simulated, including governing equations (or main model) and assumptions
\item Units, coordinate system, boundary/initial conditions (if applicable)
\item What each data field represents physically
\end{itemize}

\textbf{C. Dataset generation process}
\begin{itemize}
\item High-level pipeline overview (steps), inputs $\to$ outputs mapping, file formats, directory structure
\item Parameterization: what parameters vary and their ranges
\item Randomness: seeds, noise models, determinism/reproducibility
\item Any filtering/post-processing steps
\end{itemize}

\textbf{D. Plots (optional):}
\begin{itemize}
\item Only create plots if they materially improve understanding (e.g., distributions, example trajectories/fields).
\item Save scripts in \lstinline|/output/dataset/documentation/scripts| and plots in \lstinline|/output/dataset/documentation/plots|.
\item Embed the plots in \lstinline|documentation.md| under the relevant section(s), as inline Markdown image links using relative paths.
\end{itemize}

\textbf{Restrictions:}
\begin{itemize}
\item Inside \lstinline|/output/dataset|, you are only allowed to create and modify the folder \lstinline|/output/dataset/documentation|.
\end{itemize}

\textbf{Style:}
\begin{itemize}
\item Prefer using equations over descriptions.
\item \lstinline|documentation.md| should use clear sections and be thorough but not verbose.
\item \lstinline|summary.md| should be a short structured summary with 2--4 mini-sections (200--500 words).
\end{itemize}

\textbf{Goal:}
\begin{itemize}
\item After reading \lstinline|summary.md|, a particle physics PhD student should be able to:
\begin{itemize}
\item identify what physical process(es) the dataset represents,
\item understand the main observables/labels and their units/meaning,
\item understand the intended use-cases and key limitations at a high level.
\item generate a new dataset with the existing scripts.
\end{itemize}
\item After reading \lstinline|documentation.md|, a particle physics PhD student should be able to:
\begin{itemize}
\item reproduce the dataset generation end-to-end on a clean machine (including exact software versions and a verification/smoke test) using only \lstinline|documentation.md| (without access to the original machine environment),
\item interpret every stored field physically (definitions, units, coordinate conventions, and any transformations),
\item understand the modeling assumptions and validity regime (what is included/excluded, approximations, and known biases),
\item determine which physics questions the dataset is suitable/unsuitable for, with explicit examples and failure modes.
\end{itemize}
\end{itemize}

\vspace{1em}
\hrule
\vspace{1em}

\textbf{Prompt explanation:} We carefully specify the requested task, deliverables, including must-have requirements, restrictions, and the documentation style. In addition, we include the intended use of the documentation, which allows \ma to further tailor the documentation.
\end{userblock}
\ma develop and execute the multi-step plan
\begin{planblock}[Plan]
\begin{enumerate}[leftmargin=*, label=\textbf{Step \arabic*:}, itemsep=0.6em]
\item Inspect \lstinline|/output/dataset| structure and key files
\item Reconstruct the software environment and dependencies
\item Analyze physics setup and generation pipeline
\item Inspect \lstinline|.npz| dataset structure and map fields to physical quantities
\item Create optional plots and define verification/smoke tests
\item Write concise summary in \lstinline|summary.md|
\end{enumerate}
\end{planblock}
and produce a documentation file
\markdownfile[summary.md]{markdown_files/summary.md}
The detailed documentation is shown in App.~\ref{app:ds_documentation}, including a few highlights: In addition to listing the software versions, they list the template as well. We uninstalled uproot such that \ma cannot extract the version. In the documentation, this is explicitly stated and a version recommendation is given. \ma find that the top mass and width are (slightly) different between the Standard Model and UFO runs. The $K$-factor construction is described in detail. A table summarizes the layout of the final dataset, including data types and units.

\clearpage
\subsection{Supporting experienced users}
\label{sec:apps_adv} 

\ma do not only support users with or without basic knowledge of \mg, but also experienced users who want to perform more advanced tasks. As an example, we consider a scenario where we want to generate high-precision production events  
\begin{align}
 pp \to t \bar{t} 
 \qquad \text{(production beyond on-shell/LO, no decays)} \; .
\end{align}
We assume that the LO simulations are not sufficient and prompt \ma on more precise simulations:

\begin{userblock}[User]
I am a PhD student working on precision simulations for top-quark processes at colliders.\\
I currently have a baseline leading-order, on-shell simulation for a given process and observable.\\

\textbf{Task:}\\ 
Critically assess the limitations of such a setup and propose systematic and conceptually distinct ways to improve its theoretical precision.

For each proposed improvement:
\begin{itemize}
    \item Explain which physical effects are being captured that are missing at baseline level.
    \item Clarify when and why this improvement becomes important for realistic analyses.
    \item Indicate, at a high level, how such an improvement is typically implemented in modern Monte-Carlo simulations.
\end{itemize}

Reason from first principles and standard practice in collider phenomenology. The goal is to identify the dominant axes along which precision is improved, not to enumerate tools.
\end{userblock}
\begin{promptblock}[MadAgents (shortened)]
[\ldots]\\ 
\textbf{Dominant and conceptually distinct precision axes}\\
Starting from a leading-order, on-shell baseline for top-quark processes, the main orthogonal directions along which theoretical predictions can be systematically improved are:
\begin{enumerate}
  \item \textbf{Perturbative QCD accuracy in production and decay:}
  fixed-order calculations at NLO or NNLO, matching to parton showers (NLO+PS), and higher-order resummation at NNLO accuracy.

  \item \textbf{Description of additional hard radiation:}
  inclusion of multi-jet matrix elements and their consistent merging with parton showers at LO or NLO accuracy.

  \item \textbf{Finite-width and off-shell effects:}
  treatment beyond the narrow-width approximation, including the full resonance structure, non-resonant contributions, and interference effects.

  \item \textbf{Electroweak and QED corrections:}
  NLO electroweak effects and photon radiation, which become increasingly relevant for high-scale observables and leptonic final states.

  \item \textbf{All-order logarithmic structure:}
  higher-logarithmic resummation and improved parton-shower descriptions for observables sensitive to large scale hierarchies.

  \item \textbf{Non-perturbative, PDF, and scheme uncertainties:}
  effects from hadronisation, the underlying event, colour reconnection, parton distribution functions and $\alpha_s$, as well as choices of mass and renormalisation schemes.
\end{enumerate}

In realistic precision top-quark analyses, several of these axes are typically combined. The appropriate level of sophistication is guided by which physical effects dominate the theoretical uncertainties for the observables and phase-space regions under consideration.

\end{promptblock} 
\ma provide a list of what they consider the essential options to improve the precision, along with short explanations. We show only the final summary for the six proposed improvements, including NLO, additional jets from the matrix elements, as well as off-shell effects. Although a more experienced user has some level of familiarity with these options, they are unsure which would be suitable for their purpose. We prompt \ma to help with this decision, ask for a small number of events to be generated for each option, and a comparison of the results.

\begin{figure}[t]
    \includegraphics[width=0.495\linewidth]{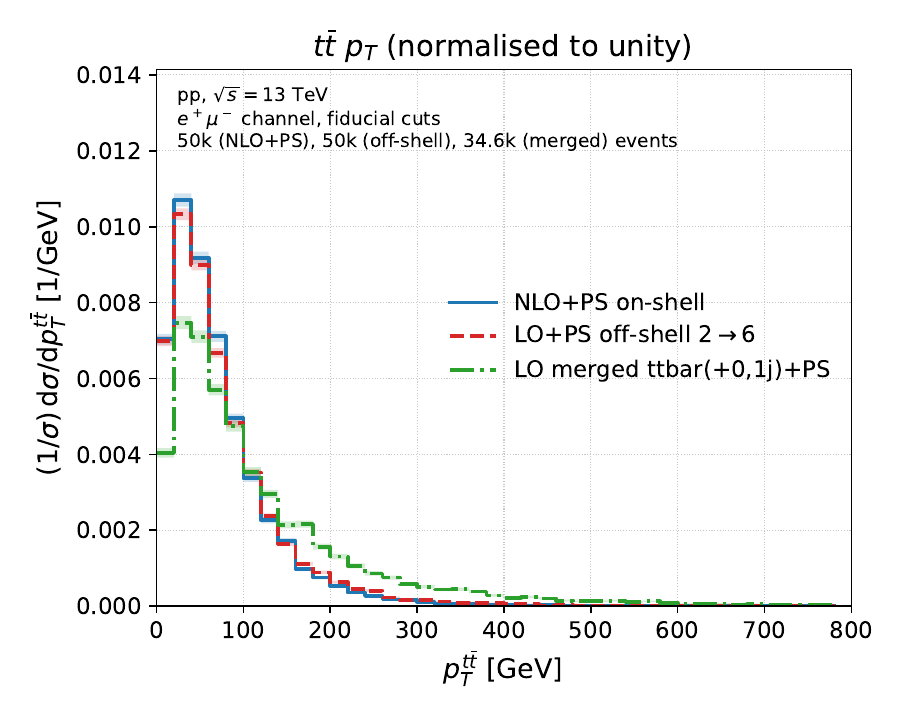}
    \includegraphics[width=0.495\linewidth]{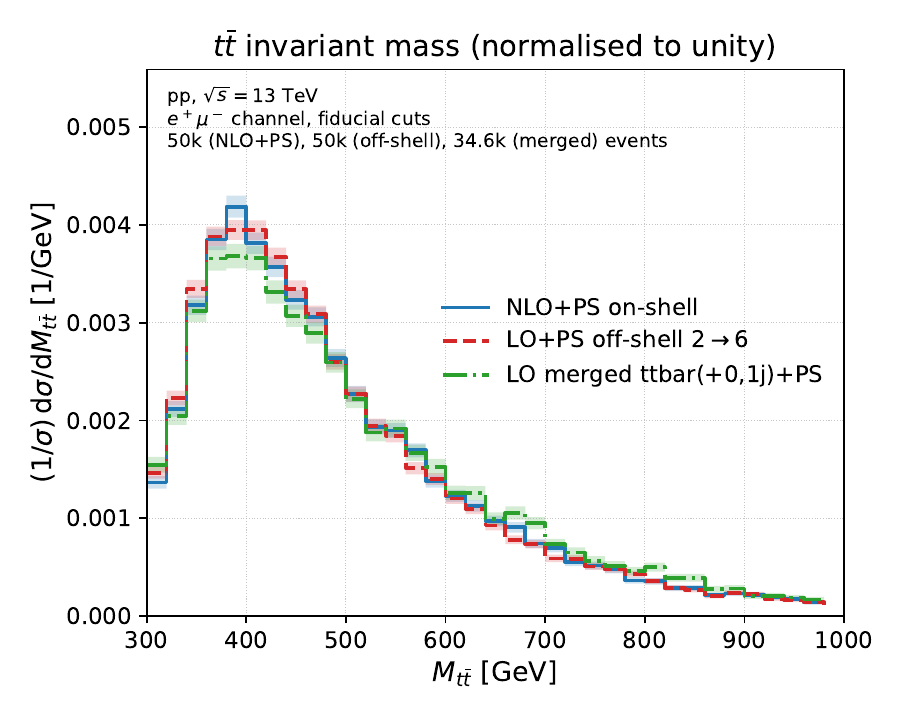}
    \caption{Comparison of the normalized $p_{T}^{t\bar{t}}$ and $m_{t\bar{t}}$ distributions for precision simulations of the production process, figure generated autonomously by \ma. Blue: NLO $t\bar{t}$ production, red: LO off-shell $2\to6$ production, green: LO $t\bar{t}$ production merged with up to one additional jet.} 
    \label{fig:ttbar_comp}
\end{figure}

\begin{userblock}[User (shortened)]
Based on the precision-improvement axes identified above, the task is to construct and compare concrete simulation setups.

\textbf{Simulation task:}
For each of the following strategies:
\begin{enumerate}
  \item Higher-order QCD corrections (NLO+PS),
  \item Finite-width and off-shell effects beyond the narrow-width approximation,
  \item Hard additional radiation via matrix-element multi-jet merging,
\end{enumerate}
do the following:
\begin{itemize}
  \item Define a representative simulation setup implementing the respective strategy.
  \item Generate 50000 events for each setup using reasonable, standard choices.
  \item Clearly state all assumptions and simplifications, such as on-shell versus off-shell decays, jet multiplicities, or matching and merging schemes.
\end{itemize}

You may choose suitable tools, schemes, and defaults where needed, but avoid unnecessary over-optimisation.
[\ldots]
\end{userblock}
\begin{promptblock}[MadAgents (shortened)]
\subsubsection*{Concrete setups and files}

\textbf{Common to all three:}
\begin{itemize}
  \item Collider: pp at $\sqrt{s}=13$ TeV.
  \item $m_t = 172.5$ GeV, $\Gamma_t \approx 1.32$ GeV.
  \item PDFs: NNPDF3.1 NLO (LHAPDF ID 303400).
  \item Final state used in the comparison: \textbf{only} $t\bar t\to b e^+\nu_e\, \bar b \mu^-\bar\nu_\mu$ (no $e^-\mu^+$).
  \item Fiducial selection (analysis level): exactly 1 e and 1 $\mu$ (OS), $p_T>25$ GeV, $|\eta|<2.5$, isolated; at least 2 jets with $p_T>25$ GeV, $|\eta|<2.5$ and at least 1 truth b-tag.
\end{itemize}

\textbf{Samples and locations:}

\begin{enumerate}
  \item \textbf{NLO+PS on-shell (higher-order QCD axis)}
  \begin{itemize}
    \item Process: $pp \to t\bar t$ [QCD] at \textbf{NLO QCD}, on-shell tops.
    \item Decays: LO MadSpin, $t\to b e^+\nu_e$, $\bar t\to \bar b\mu^-\bar\nu_\mu$.
    \item Matching: aMC@NLO to Pythia8, hadronisation+UE on. [\ldots]
  \end{itemize}

  \item \textbf{LO+PS off-shell 2$\to$6 (off-shell axis)}
  \begin{itemize}
    \item Process: $pp \to e^+\nu_e\,\mu^-\bar\nu_\mu\,b\bar b$ at \textbf{LO QCD}, with complex-mass scheme and all doubly-, singly- and non-resonant diagrams (full $W^+W^-b\bar b$ structure).
    \item No separate decay step: leptons and b’s appear directly from the ME, with finite widths.
    \item Shower: Pythia8, hadronisation+UE on. [\ldots]
  \end{itemize}

  \item \textbf{LO MLM-merged $t\bar t$(+0,1j)+PS (hard-radiation axis)}
  \begin{itemize}
    \item Processes: $pp \to t\bar t$, $pp \to t\bar t j$ at \textbf{LO QCD} (0 and 1 extra parton; 2-jet MEs were dropped due to a technical issue).
    \item Merging: MLM $k_T$ with Pythia8, $x_{\text{cut}} \approx 30$ GeV, dynamic scale $\sim H_T/4$.
    \item Decays: LO MadSpin to the same $e^+\mu^-$ channel as in NLO+PS. [\ldots]
  \end{itemize}
\end{enumerate}

\textbf{Unified analysis and plots:}
\begin{itemize}
  \item {[\ldots]}
  \item Plots (PDF+PNG): \texttt{plots/compare\_ttbar\_*.pdf} (Njets, pT\_lep1, pT\_lep2, pT\_ttbar, M\_ttbar, M\_bl, with log-y variants for Njets and pT\_ttbar).
\end{itemize}

[\ldots]\\
If you want, we can next pick one observable (e.g.\ $p_{T,t\bar t}$) and go through the plots and numbers in more detail, or adapt the selections to a concrete analysis you’re working on.
\end{promptblock}
The response by the \ma summarizes their workflow effectively. They begin by specifying parameters common among all simulations such as collider energy, final state particles and analysis level cuts.
This is both efficient and ensures comparability between simulations. 
They continue with the settings for the individual simulations. Here, \ma realizes that the produced tops need to be decayed if the on-shell simulations are to be compared with the off-shell production. For this \ma decides to implement the top decays via MadSpin~\cite{Artoisenet:2012st}.\\
Not shown in its reply but nonetheless worth mentioning is that, because we originally only provide a minimal \mg environment, the NLO simulation requires additional packages be installed. Again, \ma show that they are able to install all necessary dependencies autonomously, as part of its workflow.
Finally, without any additional user input, \ma select relevant kinematic observables, including $p_{T}^{\ell_1},p_{T}^{\ell_2},p_{T}^{t\bar{t}}$, the invariant mass $m_{t\bar{t}}$, and the number of jets $N_\text{jets}$. After running all simulations, \ma analyze the events and create histograms for these observables.
In Fig.~\ref{fig:ttbar_comp} we show the normalized transverse momentum $p_{T}^{t\bar{t}}$ and invariant mass $m_{t\bar{t}}$ distributions of the top pair created autonomously, where the normalization mixes radiative effects. We observe differences especially for the transverse momentum, but we want to know more and prompt \ma to analyze and discuss them.

\begin{userblock}[User]
Focusing on the plots you have generated, examine in detail the normalized transverse-momentum distribution of the $p_{T,t\bar{t}}$.

\textbf{Task:}
\begin{itemize}
    \item Compare the shapes of the distributions obtained with the different simulation strategies.
    \item Identify and explain the dominant physical effects responsible for the observed differences in shape.
    \item Discuss how these differences impact the interpretation of measurements or fits.
\end{itemize}
\end{userblock}

\begin{promptblock}[MadAgents (shortened)]
\subsubsection*{Qualitative comparison of shapes}

\textbf{NLO+PS on-shell vs LO+PS off-shell $2\to6$.}
\begin{itemize}
  \item Once you normalise each curve to unit area, the shapes are \textbf{very similar} over most of the $p_{T,t\bar t}$ range. [\ldots]
\end{itemize}

\textbf{LO merged $\bar t t(+0,1j)+\mathrm{PS}$ vs the above two.}
\begin{itemize}
  \item The \textbf{merged} sample is visibly \textbf{harder} in $p_{T,t\bar t}$
  [\ldots]
  \item After unit-area normalisation, the merged curve sits \textbf{below} NLO at small $p_T$ and \textbf{above} it in the tail.
\end{itemize}
[\ldots]
\vspace{1em}
\hrule

\subsubsection*{Dominant physical effects behind these differences}
\textbf{(a) Why NLO+PS on-shell and LO off-shell shapes are so similar}

\begin{itemize}
  \item In both cases, the \textbf{transverse recoil} of the $t\bar t$ system is governed primarily by \textbf{QCD radiation}: the first hard emission and the subsequent shower.
  \item The off-shell $2\to6$ sample differs from on-shell NLO mainly by its \textbf{resonance structure}:
  \begin{itemize}
    \item It includes doubly-resonant, singly-resonant, and non-resonant diagrams with finite top and $W$ widths.
  \end{itemize}
\end{itemize}
[\ldots]\\
\textbf{(b) Why the LO merged $\bar t t(+0,1j)+\mathrm{PS}$ spectrum is harder}\\
Here the key difference is not off-shellness or perturbative order per se, but the \textbf{matrix-element description of hard extra radiation}:

\begin{itemize}
  \item In the merged sample you have explicit LO matrix elements for
  \[
    pp \to t\bar t \quad \text{and} \quad pp \to t\bar t + j,
  \]
  combined with MLM matching. This means:
  \begin{itemize}
    \item The \textbf{first hard jet} is described by the \textbf{exact tree-level ME} rather than purely by the shower.
  \end{itemize}
\end{itemize}
[\ldots]

\vspace{1em}
\hrule

\subsubsection*{Impact on measurements and fits}
\textbf{(a) Measurements of $\mathrm{d}\sigma/\mathrm{d}p_{T,t\bar t}$}

For a differential $t\bar t$ measurement in $p_T$:
\begin{itemize}
  \item If you use \textbf{NLO+PS on-shell} as your baseline, you get:
  \begin{itemize}
    \item Correct inclusive normalisation (at NLO) and a decent shape at low--moderate $p_T$.
    \item Some residual modelling uncertainty in the \textbf{tail}, where multiple hard emissions are important.
  \end{itemize}

  \item If you use only \textbf{LO+PS off-shell} without merging:
  \begin{itemize}
    \item You gain a better description of off-shell line shapes, but
    \item You still suffer from LO-only radiation in production; in the $p_T$ tail, uncertainties are large and the shape can deviate from higher-order predictions.
  \end{itemize}

  \item If you use \textbf{LO merged $\bar t t(+0,1j)+\mathrm{PS}$}:
  \begin{itemize}
    \item You get a much more physical shape in the region where a single hard extra jet dominates the recoil.
    \item However, the normalisation is still LO; for serious comparisons you’d prefer NLO merged (or at least NLO+LO-merged) setups.
  \end{itemize}
\end{itemize}

[\ldots]\\
\textbf{(b) Fits to PDFs and EFT / BSM interpretations}\\
Both PDF fits and EFT/BSM searches often look for deviations in the \textbf{high-$p_T$ tail}:

\begin{itemize}
  \item If the SM prediction \textbf{underestimates} the tail (e.g.\ pure LO+PS without merging, or inadequate treatment of higher multiplicities), you might misinterpret an excess as evidence for new physics or anomalous couplings when it’s just missing hard QCD radiation.
  \item If the SM prediction \textbf{overestimates} the tail (e.g.\ some LO merged setups can overshoot in particular regions if scales/merging parameters are not well chosen), you could conversely weaken limits or miss real deviations.
\end{itemize}

[\ldots]\\
In short: for the \textbf{normalized} $p_{T,t\bar t}$ distribution, the main driver of shape differences is whether you include \textbf{matrix-element multi-jet information} (the merged sample), not whether you use off-shell vs on-shell top kinematics at fixed order. That has direct consequences for how you model high-$p_T$ tails in measurements, PDF fits, and EFT/BSM interpretations; off-shell effects then become important in more specialized, kinematic-edge or interference-dominated observables rather than in a standard, central $p_{T,t\bar t}$ spectrum.
\end{promptblock}
\ma reply with a detailed discussion of the plot. They begin by visually analyzing the differences between the methods, pointing out both similarities and differences between the approaches, while also providing a brief discussion of possible causes from physics effects.
Particularly interesting are also the physics scenarios \ma independently came up with. They begin by discussing Standard Model precision measurements, list the pros and cons of each approach and give suggestions which is best depending on their purpose. 
They recognize that the simulations mostly differ in the tails and extend the discussion to PDF and EFT fits, where the proper modeling of these tails is essential. This illustrates how \ma can help users interpret results in the context of physics scenarios they themselves might be unfamiliar with, such as PDF and EFT fits in this case. In this example, \ma run for approximately 2:30 h.

\clearpage
\subsection{Autonomous event generation}
\label{sec:apps_auto}

Motivated by the workflow suggested by HEPTAPOD~\cite{Menzo:2025cim}, we perform a scan over masses for a simple scalar leptoquark model. As input we provide \ma with a pdf file containing only the information provided in the example workflow section of their paper. This includes a brief description of the leptoquark model and its decays, as well as a rough outline of the workflow. From parton-level generation with \mg to showering and hadronization with \pythia, jet clustering with \fastjet, and the reconstruction of the minimum leptoquark mass defined as
\begin{align}
    m_{\mathrm{LQ}}^{\mathrm{min}} = \min \left\{m_{\mathrm{LQ}}^{(1)},m_{\mathrm{LQ}}^{(2)}\right\} \; .
\end{align}
We do not provide any specific information on how to perform this task, we only outline the steps and tools to be used and the following prompt. The exact file used can be found \href{https://github.com/MadGraphTeam/MadAgents/blob/main/supplementary/autonomous_generation/heptapod_input.pdf}{here}.

\begin{userblock}[User]
Please implement and execute the workflow described in the paper.
As a final deliverable, I want the generated events, a histogram of the resonance reconstruction, and a bash script that can execute the end-to-end workflow (including creation of the resonance histogram).
All deliverables must be saved, you may create suitable subdirectories.\\

\textbf{Constraint}:
Your generated events and reconstructions must match the paper as closely as possible.\\

\textbf{Persistence}:
Do not present partial or simplified results. Present the final results only if you believe you have recreated the workflow to the best of your ability and that the outputs match the paper as closely as reasonably possible.\\

\textbf{Assumptions \& Decisions}:
If any part of the paper is unclear or ambiguous, do not ask me. Instead, make a reasonable decision. In your final reply, list every unclear or ambiguous point you encountered and, for each one, state the decision you made and briefly explain your reasoning.\\

\textbf{Permission}:
You are allowed to install any missing dependencies. Do not ask me for installation decisions; make reasonable decisions yourself.

\end{userblock}
Since this is a non-trivial task \ma decides that a step-by-step plan needs to be created.

\begin{planblock}[Plan]
\begin{enumerate}[leftmargin=*, label=\textbf{Step \arabic*:}, itemsep=0.6em]

\item Locate PDF and set up project directories.

\item Extract physics and workflow specifications from the PDF and references.

\item Inspect and prepare the software environment.

\item Implement MadGraph and event generation workflow.

\item Implement reconstruction, resonance analysis, and histogram production.

\item Create end-to-end bash workflow script.

\item Validate full workflow and prepare final deliverables and ambiguity summary.

\end{enumerate}
    
\end{planblock}
For this complex task, the Orchestrator uses the Script-Operator to set up a well-defined environment to work in, to make the workflow as smooth as possible. Afterwards, the PDF-Reader searches the pdf file for the information needed, after which the Script-Operator logs ambiguities. In the next step, the MG-Operator checks if all required tools are installed, installing missing dependencies as needed. In this example, the only missing tool is the UFO file for the leptoquark model. To remedy this, \ma launch a web search to find a public UFO model.

With the setup established, \ma begin with the implementation of the \mg workflow. This includes writing the \texttt{run\_cards}, \texttt{param\_cards}, and other configuration files. They are tested and debugged through low-statistics test runs, to confirm that they generate valid events. They are further inspected by the Reviewer to ensure that they are consistent with the specifications provided in the input. Once events are generated, they need to be analyzed. The Script-Operator writes the reconstruction code, followed by the Plotter for visualization. Again, the Reviewer checks its validity. Finally, \ma create an end-to-end bash script for easy reproduction and validation of the workflow. Once this is done, all requested outputs, \eg event files, figures, and code used for the analysis, are consolidated into a single output folder.

\begin{figure}
    \centering
    \includegraphics[width=0.7\linewidth]{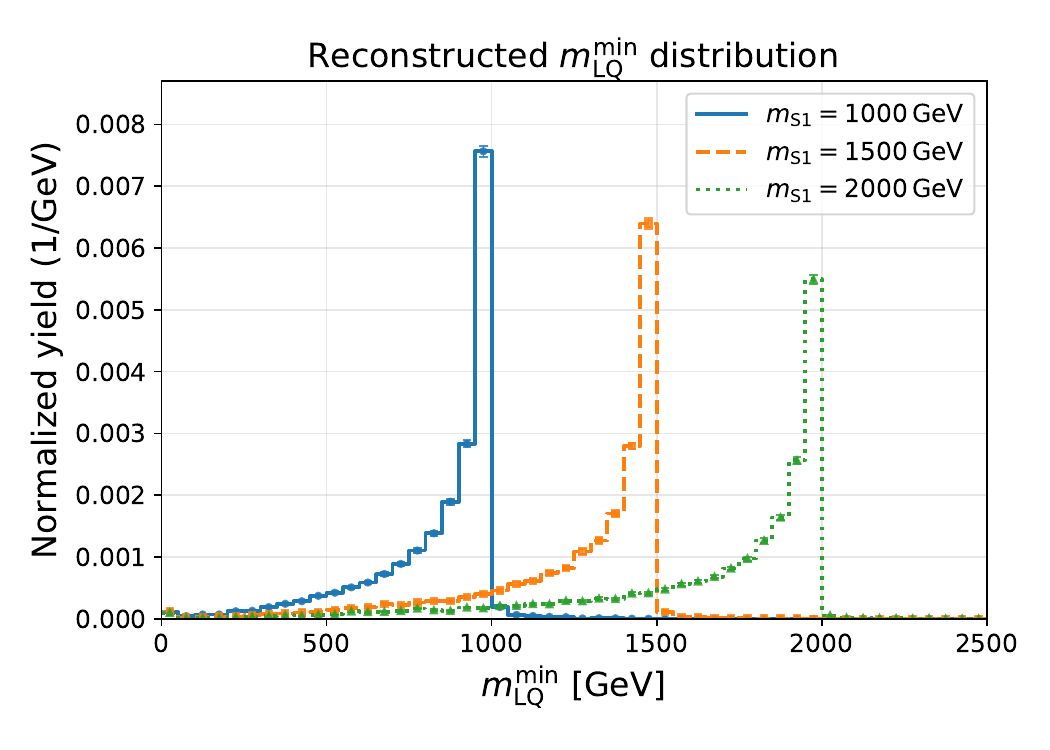}
    \caption{Reconstructed minimum leptoquark mass for $m_{S1} \in [1, 1.5, 2]$ TeV, figure generated autonomously by \ma.}
    \label{fig:HEPTAPOD_comp}
\end{figure}

Figure~\ref{fig:HEPTAPOD_comp} shows the minimum reconstructed invariant mass of the leptoquark, created completely autonomously by the Plotter. We can clearly see the  resonance peaks for each simulated leptoquark mass. Although this simulation chain is executed autonomously, \ma are not meant to replace the human user. It should support and teach the user how to complete these tasks. To this end, we prompt \ma to give details on the essential steps. \ma not only return a concise summary, but also point the user to the scripts and notes it created for more details. That way, the user is provided with all information needed to understand the simulations, validate them and implement adjustments as needed.

\begin{userblock}[User]
    Please generate a clear, well-structured summary of the most important steps you carried out and the key decisions you made, including appropriate citations, written in a style suitable for direct inclusion in a LaTeX document.
\end{userblock}

\begin{promptblock}[MadAgents (shortened)]
\textbf{Simulation and analysis setup}\\
The signal under study is a simplified scalar leptoquark $S_1$ with quantum numbers
$S_1 \sim (\bar{3},1,1/3)$, pair-produced in proton--proton collisions,
followed by the decay chain
\begin{align}
  pp \to S_1 S_1^{\dagger} \to (\ell^+ j)(\ell^- j)
  \;
  \Rightarrow \;
  2\ell + 2j~\text{final state},
\end{align}
closely following the workflow outlined in the HEPTAPOD~\cite{Menzo:2025cim} short note and the
MC4BSM-style tutorials.\\

\textbf{Event generation.}\\
Events are generated with \texttt{MadGraph5\_aMC@NLO} v3.7.0~\cite{Alwall:2014hca}
interfaced to \texttt{Pythia~8.3}~\cite{Sjostrand:2014zea} for parton showering and
hadronisation. We use an existing public leptoquark UFO model,
\texttt{LQnlo\_5FNS\_v5\_UFO}, originally constructed in the
\texttt{FeynRules} framework~\cite{Alloul:2013bka}, and identify the HEPTAPOD
scalar $S_1$ with the particle \texttt{LQ1d} (and its antiparticle
\texttt{LQ1d\textasciitilde{}}). The collider setup is
\begin{itemize}
  \item proton--proton collisions at $\sqrt{s} = 8~\text{TeV}$, implemented as
        $E_{\text{beam}1} = E_{\text{beam}2} = 4000~\text{GeV}$;
  \item leading-order (LO) matrix elements with parton shower (LO+PS),
        no ME--PS matching;
  \item PDFs: LHAPDF6~\cite{Buckley:2014ana} with the NNPDF3.1 LO set
        NNPDF31\_lo\_as\_0130~\cite{Ball:2017nwa} (\texttt{lhaid} = 315200).
\end{itemize}

The leptoquark mass is scanned over three benchmark points
\begin{align}
  m_{S_1} \in \{1.0, 1.5, 2.0\}~\text{TeV}.
\end{align}
For each mass point, we generate $N_{\text{events}} = 20\,000$ unweighted LO
parton-level events and shower them with Pythia~8.  The total width of $S_1$ is
taken from the UFO model defaults (narrow-width regime, $\Gamma/m \lesssim 1\%$).
A simple flavour structure is chosen by turning on only the right-handed
Yukawa coupling between first-generation up quarks and electrons,
\begin{align}
  y_{11} \ne 0,\quad y_{ij} = 0~\text{for}~(i,j) \neq (1,1),
\end{align}
so that $S_1$ decays dominantly as $S_1 \to e^+ u$ and $S_1^{\dagger} \to e^- \bar u$.
SM backgrounds (DY+jets, $t\bar t$, dibosons) are not simulated in this
implementation; the focus is on the signal resonance shape and its dependence on
$m_{S_1}$.

\textbf{Hadron-level reconstruction.}\\
The analysis is performed at hadron level, using Pythia~8 final states without
additional detector simulation.  Stable final-state particles are taken as those
with \texttt{status} $=1$ in the HepMC record~\cite{Dobbs:2001ck}. Neutrinos
($\lvert \text{PID} \rvert \in \{12,14,16\}$) are removed from both lepton and
jet reconstruction.

Electrons are reconstructed from HepMC particles with $\lvert \text{PID} \rvert = 11$,
with kinematic cuts
\begin{align}
  p_T^e > 50~\text{GeV},\qquad \lvert \eta^e \rvert < 2.5,
\end{align}
and we require exactly two such electrons per event.

Jets are clustered from all remaining stable particles (excluding neutrinos and
electrons) using the anti-$k_T$ algorithm~\cite{Cacciari:2008gp} with radius
parameter $R=0.4$ as implemented in the \texttt{FastJet} framework
(via the Python wrapper \texttt{pyjet})~\cite{Cacciari:2011ma}. Reconstructed jets
must satisfy
\begin{align}
  p_T^j > 50~\text{GeV},\qquad \lvert \eta^j \rvert < 5.0,
\end{align}
and at least two jets are required. Only the two leading jets in $p_T$ are used
for the resonance reconstruction.

\textbf{Resonance reconstruction and $m_{LQ}^{\min}$.}\\
In events with exactly two selected electrons $(\ell_1, \ell_2)$ and at least
two selected jets $(j_1, j_2)$, we reconstruct two leptoquark candidates by
forming all possible pairings between leptons and the two leading jets:
\begin{align}
  \text{Pairing A}: &\quad (\ell_1 + j_1),~(\ell_2 + j_2),\\
  \text{Pairing B}: &\quad (\ell_1 + j_2),~(\ell_2 + j_1).
\end{align}
For each pairing we compute the invariant masses $(m_1, m_2)$ of the two
candidates. We select the pairing that minimises $\lvert m_1 - m_2 \rvert$, and
from that pairing define the reconstructed leptoquark mass variable
\begin{align}
  m_\text{LQ}^{\min} = \min(m_1, m_2).
\end{align}
This follows the standard HEPTAPOD/MC4BSM prescription and produces a
resonance-like peak near the true $m_{S_1}$ with a modest low-mass tail. For a
broader discussion of scalar leptoquark phenomenology and typical benchmark
choices, see e.g. Ref.~\cite{Dorsner:2016wpm}.

\textbf{Histogramming and normalisation.}\\
The following binning of $m_{LQ}^{\min}$ is used:
50 bins from $0$ to $2500~\text{GeV}$ with unit-area normalisation per mass point, i.e. each distribution is normalised such that $\sum_i N_i = 1$. The bin contents can be interpreted as probabilities per bin (or, equivalently, as a discretised probability density when divided by the bin width). We overlay the three benchmark spectra ($m_{S_1}=1.0,1.5,2.0$~TeV) for direct comparison. Poisson uncertainties $\sqrt{N_i}$ are propagated to the normalised histograms and shown as error bars.

[\ldots]\\
All non-trivial choices that are not fixed by the HEPTAPOD note (collider
energy, PDF set, specific UFO model, Yukawa pattern, use of hadron-level
rather than detector-level objects, exact cuts and binning) are documented in a
separate ``ambiguities and decisions'' note that accompanies this analysis.
\end{promptblock}

For this example, the final results of the autonomous campaign are obtained after approximately 3:30 h.

\clearpage
\section{Outlook}
\label{sec:outlook}

Modern ML methods allow us to combine the two challenges of LHC physics, the vast amount of data and its complexity, into a unique opportunity. First-principle theory simulations in the form of event generators allow us to formulate our understanding of the LHC in terms of a fundamental physics Lagrangian. If we want to use ML-enhanced simulation based inference to extract all available physics from the LHC dataset, the numerical tools and the analysis workflow both have to become more precise, faster, and more efficient. Agentic use of ultra-fast simulation tools is re-shaping numerical science, so we have explored accelerated event generation, specifically a highly efficient user interface for \mg using \ma:
\begin{enumerate}
\item First, we have shown how \ma can autonomously install the \mg software, even including building a \root from the source code.
\item Second, we have illustrated how \ma can communicate with inexperienced users to provide tutorials and support for standard tasks. The user can actively interact with \ma and ask for explanations, for instance, how to re\-weight SMEFT events with $K$-factors computed from the SM.
\item Next, we have demonstrated how \ma provide support to experienced users. They generated events employing non-trivial theory approaches to top pair production ---  NLO production, off-shell tops, and additional matrix element jets --- and analyze the differences in detail.
\item Finally, we have turned to an autonomous task, namely, independently reproducing simulations from a publication. We have shown how \ma reproduce the key figures of the HEPTAPOD publication, including a detailed account of all relevant pieces of information missing from the publication.
\end{enumerate}
This study has shown how agentic support of tools like \mg will significantly accelerate LHC research. Autonomous simulations provide an important step towards AI-accelerating not only LHC searches, but also precision analyses and global analyses. In addition, we have seen how \ma take over parts of our workflows and at the same time train physicists.

\section*{Acknowledgments}

We would like to thank Siddharth Mishra-Sharma for providing us with free Anthropic Claude access and for triggering the new Claude Code implementation described in App.~\ref{sec:claude} of this updated paper. We would also like to thank Theresa Reisch, Giovanni Ottaviano, and Tobias Golling for their encouragement, support, and advice through this proof-of-principle study. We are also grateful to Olivier Mattelear for pointing out the wrong statement in discussing off-shell Higgs decays. In addition, TP would like to thank Michael Kagan for a huge number of inspiring discussions about future workflows in particle physics. This research is supported by the Deutsche Forschungsgemeinschaft (DFG, German Research Foundation) under grant 396021762--TRR~257: \textsl{Particle Physics Phenomenology after the Higgs Discovery}, and through Germany's Excellence Strategy EXC~2181/1 -- 390900948 (the \textsl{Heidelberg STRUCTURES Excellence Cluster}). We acknowledge support by the state of Baden-Württemberg through bwHPC and the German Research Foundation (DFG) through grant no INST 39/963-1 FUGG (bwForCluster NEMO). The exchange with the Geneva group is funded by the European Union’s Horizon Europe research and innovation programme Cofund SOUND.AI under the Marie Sklodowska-Curie Grant Agreement No 101081674. Views and opinions expressed are however those of the author(s) only and do not necessarily reflect those of the European Union or the granting authority. Neither the European Union nor the granting authority can be held responsible for them.

\clearpage
\appendix
\section{Individual agents} 
\label{app:agentic_details}

\appsubsection{Controlling agents}

The Orchestrator, Planner and Reviewer share the same conversation history. Upon invocation, each agent first generates reasoning tokens and then outputs its response. Since the reasoning tokens are agent and system prompt specific, we do not insert them into the shared conversation. While there is the option to include the reasoning tokens belonging to the invoked agent on the fly, we have decided against it as we observed that they confuse the Orchestrator: Instead of proceeding with the workflow, the Orchestrator repeatedly called the same agent with almost identical instructions.

As the Reviewer is able to call tools (web-search does not count as tool calls in this sense as we use the \href{https://platform.openai.com/docs/guides/tools-web-search}{OpenAI's built-in tool}), we create a temporary copy of the shared conversation and append all tool calls, tool outputs and reasoning tokens. The latter is explicitly \href{https://platform.openai.com/docs/guides/reasoning#keeping-reasoning-items-in-context}{recommended by OpenAI for tool-calling agents}. After the Reviewer responds, we append only the answer to the shared conversation.

In case the shared conversation becomes too long, those three agents also share a common conversation summary. Again, there is a special case for the Reviewer. If we need to summarize the conversation during its tool calling, the Reviewer obtains a dedicated, temporary conversation summary. Once the Reviewer invocation finishes, a shared summary is created.

As the Plan-Updater simply translates the Orchestrator's requests into structured data, it only observes the Orchestrator's immediate instructions.

We restrict the Orchestrator, Planner and Plan-Updater to generate \href{https://platform.openai.com/docs/guides/structured-outputs?utm_source=chatgpt.com}{structured output} and refer to Ref.~\cite{geng2023grammar} for a conceptual introduction. In the following, we present their output schemas and describe how they are appended to the shared conversation.

\paragraph{Orchestrator Output}

The Orchestrator outputs its decisions in the following structured form:
\begin{itemize}
    \item \textbf{recipient:} the recipient that the Orchestrator messages/invokes next.
    \item \textbf{reasoning:} a brief explanation for the Orchestrator's decision.
    \item \textbf{message:} the message sent to the recipient. For workers: typically the task instruction.
    \item \textbf{reasoning effort:} the reasoning effort of the invoked agent.\\
    Possible values: "low", "medium", "high".\\
    This allows the Orchestrator to manage the reasoning time/cost depending on the task. The Orchestrator is always invoked with "high" reasoning effort.
    \item \textbf{future note:} a small scratchpad for near-future steps.\\
    This enables the orchestrator to memorize its upcoming steps. We briefly explain why this is useful:\\
    It is not uncommon for the orchestrator to plan a few steps ahead. Once the Orchestrator invokes an agent or responds to the user, its reasoning tokens are stripped, which might contain its short-term plan. Without this scratchpad, the Orchestrator is forced to rethink the upcoming steps every time. Not only does this increase the cost and time, it might even degrade the Orchestrator's performance.\\
    This explanation is similar to OpenAI's explanation for \href{https://platform.openai.com/docs/guides/reasoning#keeping-reasoning-items-in-context}{keeping the reasoning tokens during tool calling}.
\end{itemize}
The raw Orchestrator output (without reasoning tokens) is inserted into the shared conversation.

\paragraph{Planner and Plan-Updater Output}

The Planner generates a plan as a list of plan steps. Each plan step contains the following fields:
\begin{itemize}
    \item \textbf{ID:} a unique identifier.
    \item \textbf{title:} a concise title.
    \item \textbf{description:} of what should be done in this step.
    \item \textbf{rationale:} a brief explanation of why this step exists.
    \item \textbf{depends on:} a list of plan step IDs on which this step depends.
    \item \textbf{status:} of this step.\\
    Possible values: "pending" (can be worked on), "in progress" (currently worked on), "done" (completed), "failed", "skipped", "blocked" (a dependency must be completed or skipped first).
    \item \textbf{outcome:} of the plan step.\\
    If completed, it contains a summary of the results.\\
    If failed, it contains the error(s).\\
    If skipped, it contains an explanation for skipping this step.
\end{itemize}

To update the existing plan, the Plan-Updater outputs a list of updates consisting of the following fields:
\begin{itemize}
    \item \textbf{ID: } of the updated step.
    \item \textbf{status: } the updated status.
    \item \textbf{outcome: } the updated outcome. It can be empty.
\end{itemize}
The status and outcome of the corresponding plan steps are overwritten with the updated values.

After creating the plan and after each plan update, we programmatically update the statuses "pending" and "blocked". First, we set every "pending" step to blocked. Then, we inspect each "blocked" step. If all of its dependencies are completed or skipped, we set this plan step to "pending".

The generated plan and the updated plan are appended to the shared conversation with the following messages
\begin{planblock}[Planner message]
I have created the following plan:\\
\{\lstinline|json.dumps(plan, indent=2)|\}
\end{planblock}
\begin{planblock}[Plan-Updater message]
The updated plan is:\\
\{\lstinline|json.dumps(plan, indent=2)|\}
\end{planblock}

\appsubsection{Worker agents}

As each Worker only observes its own dedicated conversation, we keep all reasoning tokens, tool calls and tool outputs. This gives the Worker the ability to recall previous chain-of-thoughts and tool interactions (unless they were summarized). This can be useful, for instance, if the Orchestrator instructs the Worker to change a deliverable slightly, if the Worker asks for help to solve an error, or if the user asks for insights into the Worker's reasoning.

To leverage that LLMs are well finetuned for user conversations, we disguise the Orchestrator messages as user messages, giving each Worker the impression that they communicate with a user.

\appsubsection{Summarizer}

Before invoking any agent (except for the Plan-Updater) and after any tool call (except for \href{https://platform.openai.com/docs/guides/tools-web-search}{OpenAI's built-in web-search tool}), we approximate the token length of the agent's conversation. For each message, we determine the token count from OpenAI's returned usage metadata whenever possible. Otherwise, we resort to a character-length-based approximation for text and a byte-length-based approximation for images and PDF files.

If the conversation exceeds $150$k tokens, we summarize the oldest messages. We keep the last 10 non-tool-output messages and, if needed, include additional messages until (i) the trimmed conversation contains at least $10$k tokens and (ii) every tool call in the trimmed conversation has its corresponding tool output included as well. The summary is appended to the agent's instructions.

We place conversation to be summarized into the Summarizer's message history
and instruct it to view the message history as data. In this way, the Summarizer is able to observe images or PDF-files. If a previous conversation summary exists, it is appended to the Summarizer's instructions.

\appsubsection{Tools}

As some tools are shared between different agents, we list all available tools and indicate in Tab.~\ref{tab:agent_tools} which agent has access to which tool. The Orchestrator, the Plan-Updater and the Summarizer are omitted from this list as they do not have access to any tool.
\begin{itemize}
    \item \textbf{web\_search:} We use \href{https://platform.openai.com/docs/guides/tools-web-search}{OpenAI's built-in web-search tool}.
    \item \textbf{bash(commands: str):} Executes the command string and captures the stdout and stderr output. If an output stream exceeds 40kB, it is forwarded to a file. If the command execution exceeds 10 minutes, both output streams are forwarded to separate files and the agent is given the PID of the process. This gives the agent the ability to work during long-lasting executions and to terminate a command if it is stuck.
    \item \textbf{wait(minutes: float):} Puts the agent to sleep. This tool is useful for long-lasting bash commands.
    \item \textbf{apply\_patch(operations: List[ApplyPatchOp]):} Applies patch operations to files. An patch operation is defined by its type ("create\_file", "update\_file", "delete\_file"), the file path and the V4A diff string (empty for deleting a file).\\
    Our tool is intentionally modeled after \href{https://platform.openai.com/docs/guides/tools-apply-patch}{OpenAI's built-in apply\_patch tool}. This reference provides the relevant details behind the V4A diff string syntax.
    \item \textbf{read\_pdf(pdf\_file\_path: str):} Encodes a PDF file in base64 and wraps it in a file content block. The PDF file is then \href{https://platform.openai.com/docs/guides/pdf-files}{processed internally by OpenAI}.
    \item \textbf{read\_image(image\_file\_path: str):} Encodes an image in base64 and wraps it in an image content block. The image is then \href{https://platform.openai.com/docs/guides/images-vision}{processed internally by OpenAI}.
    \item \textbf{int-CLI-tools:} Collection of 4 tools used for an interactive CLI session.\\
    - int\_cli\_status() summarizes the state of the CLI. In particular, it indicates whether the agent has inspected the session before.\\
    - read\_int\_cli\_transcript(start\_line: int, end\_line: int) reads a slice of the CLI's transcript.\\
    - read\_int\_cli\_output(wait\_s: float) reads any new CLI output after waiting wait\_s seconds. If new output is received, it waits for 2 seconds and repeats until no additional output appears.\\
    - run\_int\_cli\_command(command: str, wait\_s: float): executes the command and reads the output with the read\_int\_cli\_output tool.
    \item \textbf{save\_answer(file\_path: str, content: str):} Saves the content to the file. If the file exists, it is overwritten.
    \item \textbf{list\_pdfs(search\_string: str):} Lists all PDF files matching the ".pdf"-appended search string by executing \lstinline|glob.glob(f"{search_string}.pdf")|.
\end{itemize}

\begin{table}[t]
\centering
\begin{small}
\begin{tabular}{l|cccccccc}
\toprule
Tool & Planner & Reviewer & \makecell{MG-\\Operator} & \makecell{Script-\\Operator} & \makecell{CLI-\\Operator} & Plotter & \makecell{PDF-\\Reader} & Researcher \\
\midrule
web\_search & $\checkmark$ & $\checkmark$ & $\checkmark$ & $\checkmark$ & $\checkmark$ & $\checkmark$ & $\checkmark$ & $\checkmark$ \\
\midrule
bash & & $\checkmark$ & $\checkmark$ & $\checkmark$ & $\checkmark$ & $\checkmark$ & & \\
\midrule
wait & & $\checkmark$ & $\checkmark$ & $\checkmark$ & $\checkmark$ & $\checkmark$ & & \\
\midrule
apply\_patch & & $\checkmark$ & $\checkmark$ & $\checkmark$ & & $\checkmark$ & & \\
\midrule
read\_pdf & & $\checkmark$ & $\checkmark$ & $\checkmark$ & & $\checkmark$ & $\checkmark$ & \\
\midrule
read\_image & & $\checkmark$ & $\checkmark$ & $\checkmark$ & & $\checkmark$ & & \\
\midrule
int-CLI-tools & & & $\checkmark$ & & $\checkmark$ & & & \\
\midrule
save\_answer & & & & & & & $\checkmark$ & $\checkmark$\\
\midrule
list\_pdfs & & & & & & & $\checkmark$ &\\
\bottomrule
\end{tabular}
\end{small}
\caption{Tool list for each agent. The Orchestrator and the Plan-Updater do not have access to any tool.}
\label{tab:agent_tools}
\end{table}

\appsubsection{Prompts} 
\label{app:prompts}

We tailor the prompts to OpenAI's GPT-5 models, following the \href{https://cookbook.openai.com/examples/gpt-5/gpt-5_prompting_guide}{GPT-5 prompting Guide}, the \href{https://cookbook.openai.com/examples/gpt-5/gpt-5-1_prompting_guide}{GPT-5.1 Prompting Guide} and the \href{https://cookbook.openai.com/examples/gpt-5/gpt-5-2_prompting_guide}{GPT-5.2 Prompting Guide}. In particular, we use XML tags to structure the instructions and adapt some recommended instruction blocks. We further refine the prompts with \href{https://chatgpt.com/}{ChatGPT}. The complete prompts set is available in the \href{https://github.com/MadGraphTeam/MadAgents/tree/main/supplementary/prompts}{code repository}.

\clearpage
\section{Example tutorial} 
\label{app:ex_tutorial}

\markdownfile[05\_bsm\_eft\_models/README.md]{markdown_files/tutorial_example.md}

\clearpage
\section{MadAgents improvements}
\label{app:improvements}

In this Appendix, we collect the improvements of the \ma setup developed after the demonstrations of Sec.~\ref{sec:apps}. We first address a concrete failure mode observed in Sec.~\ref{sec:apps_train_learn}, where \ma present a wrong \mg claim. In a second step, we describe a refined agent setup.

\subsection*{Avoiding incorrect MadGraph statements} 
\label{app:madgaph_mistakes}

First, we discuss \ma' incorrect statement from Sec.~\ref{sec:apps_train_learn}, where we generate the process
\begin{bashblock}
    generate p p > z h, (h > z z, z > e+ e-)
\end{bashblock}
and \ma wrongly state that only a single $Z$ boson decays. In fact, \mg decays both $Z$ bosons from the Higgs decay, as the \mg implementation requires for identical particles. The \ma explanation is plausible but wrong, because they generate this response without consulting relevant documentation or inspecting the \mg behavior or source code. This suggests that convention-dependent choices of \mg may not be inferred reliably solely by the LLM weights.

To avoid this problem, we inspect the \ma traces to understand how they arrive at the wrong statement, and iteratively refine their instructions. As a result, we add the option that explanations about \mg and related tools require evidence by adding the following two bullet points to the Orchestrator's System Prompt
\begin{promptblocklitlarge}[Orchestrator Evidence Instruction (System Prompt)]
- Never base your answers on your general knowledge of MadGraph or related tools; worker agents must present supporting evidence.
- MadGraph evidence requirement: When discussing MadGraph or related software, do not state a factual claim unless it is supported by verifiable evidence.
  Acceptable evidence includes (non-exhaustive):
  (a) an excerpt from official/trustworthy documentation,
  (b) a local software invocation (exact command + exact output), or
  (c) a relevant excerpt from the software source code (file path + snippet).
  Evidence must directly support the specific claim. Do not rely on common sense or generic/loosely related references.
  If you cannot obtain evidence, label the statement as UNVERIFIED and present it as a hypothesis, along with a concrete step to verify it (e.g., a command to run, a file/function to inspect, or a doc section to check).
  It is crucial that you NEVER present false claims to the user.
\end{promptblocklitlarge}
This way, \ma typically consult the web, invoke local \mg runs or inspect the source code. Unfortunately, this does not solve the problem completely as \ma sometimes misinterpret the \mg output or web content. Although the former can typically be spotted by the Reviewer, the latter poses a serious problem when \ma find online \mg examples to support its incorrect claim. In that case \ma report that they find evidence and confidently present the wrong statement.

As a solution, we instruct the Reviewer to accept only authoritative and explicit evidence, and especially exclude examples. We add to its System Prompt
\begin{promptblocklitlarge}[Reviewer Evidence Instruction (System Prompt)]
- Never accept a MadGraph (or associated software) related claim just because the reasoning is coherent or it is based on common sense.
  Do not extrapolate from evidence: ONLY treat a claim as verified if the evidence proofs the exact fact, even if it seems plausible.
- Never base your review on general knowledge of MadGraph or related tools; evaluate only what is supported by evidence, and you may seek additional evidence (e.g., official docs, source code, or local invocations) provided you cite it explicitly.
\end{promptblocklitlarge}
and to its Developer Prompt
\begin{promptblocklitlarge}[Reviewer Evidence Instruction (Developer Prompt)]
- When the user request involves a discussion/explanation of MadGraph or related tools, treat EVERY declarative statement as a CLAIM. Each CLAIM must be labeled VERIFIED (with evidence below) or UNVERIFIED (and not presented as fact). Reasoning/logic (e.g., “this only makes sense if…”) is NOT evidence and can NEVER justify VERIFIED.
- Evidence requirements (strict, for ALL claims):
  VERIFIED is allowed ONLY with verbatim AUTHORITATIVE evidence that directly supports the exact claim in the same MG context (version/mode/model/options):
  (a) MG5/MadGraph help output (exact command + exact output + version/mode),
  (b) official docs/manual (URL or local file path + short excerpt),
  (c) source code (file path + snippet),
  (d) minimal reproducible local test (exact commands + exact outputs).
  Everything else, including forums/Q&A (e.g. Launchpad), tutorials, blog posts, and ALL “examples” (even “official/public examples”), is NON-AUTHORITATIVE: it may be cited only to motivate where to look, but it CANNOT support VERIFIED. If the only cited support is NON-AUTHORITATIVE, the claim MUST be UNVERIFIED.

  Quality + anti-loophole check: Evidence must be exact-context and unambiguous. Evidence that something is recommended, typical, or shown in examples does NOT establish necessity or exclusivity; do not infer “only/always/never” unless authoritative evidence explicitly says so.
  Before labeling any claim VERIFIED, do a quick adversarial self-check: “If I remove my intuition/pattern-matching/analogy, does the quoted authoritative evidence STILL force the claim to be true?” If not, the claim is UNVERIFIED. If ambiguity remains, resolve via (a)/(c)/(d) or keep UNVERIFIED.

  Two-stage rule:
  1) First, judge using ONLY evidence already provided.
  2) For remaining UNVERIFIED claims, attempt verification only if the claim is material; use only (a)-(d) and show what you checked. Otherwise keep UNVERIFIED and recommend drop/reword as explicitly UNVERIFIED.
- If you label any claim as VERIFIED in your final answer, you MUST include: "Evidence type: (a)/(b)/(c)/(d)" and the corresponding verbatim excerpt/output/snippet. Otherwise the claim MUST be labeled UNVERIFIED and must not be presented as fact.
\end{promptblocklitlarge}
This further reduces the risk of a factual misstatement.

\subsection*{Refined agent setup}

Beyond the above presented failure mode, we improve the agent setup further. We extend the set of agents by adding a Physics-Expert worker, instructed to provide high-quality physics reasoning, and by splitting the Reviewer into Plan-Reviewer, Verification-Re\-viewer, and Presentation-Reviewer. They are tasked with planning, correctness, and presentation, and they detect more errors~\cite{d2024marg}. Finally, we replace the Plan-Updater agent by a plan updating tool which avoids an unnecessary LLM invocation. 

We separate the shared conversation between the Orchestrator, Planner, and reviewing agents into agent-specific conversations, as LLMs are typically trained on single agent conversations. To provide user context to the Planner and the reviewing agents, we insert the user-orchestrator conversation inline into their queries. When invoking workers the Orchestrator can now select a plan step which inserts a description of the corresponding plan step into the worker query. This gives worker agents greater awareness of their position in the workflow.

Next, we allow the Planner to explore the current environment before finalizing a plan. This way, the plan does not require exploration steps and can be tailored to the current setup~\cite{song2023llm}. Furthermore, the plan is no longer created via structured output. Instead, we equip the Planner with plan creation and modification tools, which avoid forcing it to provide a correct plan at once. It can create the plan step by step and iterate on the draft, for example to fix typos.

Finally, the Orchestrator can now issue tool calls, preserving the chain of thought, and  dispatch multiple agents in parallel. Each agent type, except for the Orchestrator, may have multiple instances, allowing, for example, two Researcher agents to run in parallel. Agent context can be managed flexibly, like re-using existing context or starting with a fresh context. The Orchestrator is provided with a scratchpad to write short-term notes~\cite{nye2021show}.

\clearpage
\section{Dataset documentation} 
\label{app:ds_documentation}

\markdownfile[documentation.md]{markdown_files/documentation.md}

\clearpage
\section{Claude Code with self-improvement}
\label{sec:claude}

We implement the refined \ma setup of App.~\ref{app:improvements} in Claude Code. To improve response efficiency and factual reliability, we build on approaches that learn from recorded failures and inefficiencies~\cite{shinn2023reflexion, wang2023voyager, zhao2024expel}. To this end, we equip the MG-Operator with internal \mg documentation, which we curate through a self-improvement loop with dedicated verification and review stages.

\subsubsection*{Claude Code Implementation}

We use Claude Code's built-in orchestration architecture. The main Claude Code agent naturally serves as the Orchestrator, dispatching work agents. To define the \ma Orchestrator, we append the delegation, planning, and reviewing instructions to the system prompt. We define the remaining \ma as subagents, except for the Planner and Plan-Reviewer, for which we use the built-in Claude Code functionality.

\subsubsection*{Documentation self-improvement}

We implement a self-improvement loop in our new Claude Code version. This loop tests the current understanding of \ma with respect to specific \mg topics, diagnoses issues in the internal documentation, and then improves it. It is split into six phases:
\begin{enumerate}
    \item \textbf{Question generation} \\
    An agent generates a set of questions, for which we can optionally specify a focus, \eg on NLO computations, additional requirements such as expert knowledge, and whether they should be distinct from existing questions. The agent is instructed to search the web for problems that users struggle with and base the generated questions on them. Each generated question must be accompanied by a reference answer.

    \item \textbf{Evaluating \ma} \\
    Each question is asked to a separate \ma instance. Optionally, a supervision agent checks that the response is complete and can request follow-up questions.

    \item \textbf{Extracting and validating claims} \\
    Each answer is decomposed into a list of self-contained, verifiable claims and assigned to a dedicated \ma instance, which verifies each claim. Only explicit \mg invocations, source code inspections, and physics derivations count as evidence. Every verdict must be accompanied by raw evidence and an explanation. A claim can be marked as unverifiable if \ma are confident they are unable to verify it. We accelerate this expensive phase by maintaining a database of claims and their verification verdicts. 

    \item \textbf{Grading and diagnosing} \\
    A grading agent assigns a verdict of correct, incorrect, or inconclusive to each answer. A question can be labeled as correctly answered even if the answer contains mistakes. Additionally, the grader can assign the following three auxiliary tags to the answer: indicating that the answer contains mistakes, indicating that the \ma' workflow is inefficient, for instance because the documentation does not cover a specific topic, and indicating that a reviewer flagged an issue within the \ma invocation. For incorrectly answered questions and questions with such a tag, an agent diagnoses the documentation for gaps, mistakes, and ambiguities.
    
    \item \textbf{Documentation improvement} \\
    These diagnoses are passed to a documentation improvement agent, tasked with improving the documentation to generalize to similar cases. The proposed documentation changes are reviewed by a style reviewer, ensuring a consistent documentation style, a quality reviewer, checking whether the changes are sufficiently general and appropriately placed, and a verification agent, which performs an analogue of Phase~3 on the modified documentation. If a reviewer flags an issue, the improvement agent revises the changes.
    
    \item \textbf{Re-evaluation and optional iterations} \\
    The documentation improvements are tested on problematic questions. If \ma' answer continues to exhibit issues, the documentation is refined by iterating Phases~2-6.
\end{enumerate}
Our self-improvement is based on the strong reasoning capabilities of modern LLMs. Agents can answer highly complex questions through reasoning and thorough exploration. Since this workflow is costly, agents may skip parts of the process, which can result in false claims. The verification phase forces the agent to perform a costly investigation, reducing the error rate.

To obtain a broad coverage of the \mg documentation, we run the loop repeatedly using Opus 4, varying the focus of the question generation. The resulting documentation covers 27 topics, from process syntax and parameter cards to NLO matching and merging, MadSpin decays, and EFT models. The improved documentation simplifies the workflow and guides \ma to relevant information, increasing their response speed and accuracy. We demonstrate the impact of the documentation improvement loop on two questions covering an NLO event production workflow and spin-correlated decays.

\paragraph*{NLO Gridpack Workflow}

\mg provides a gridpack mechanism that deploys precomputed integration grids across computing clusters, allowing parallel event generation without repeating the expensive grid training. At NLO this mechanism is not automated, and the equivalent workflow has to be assembled by hand from the grid-setup and event-generation steps. When asked how to create an NLO gridpack for $pp \to t\bar{t}$ production, \ma arrive at the correct procedure by inspecting the \mg source code. Their answer is correct, but the MG-Operator worked for 13:34~min, the majority of which is spent reverse-engineering the workflow from source code and the web. After adding a brief description of the procedure to the documentation, the MG-Operator no longer reconstructs the workflow and only confirms the individual commands against the source, arriving at the correct procedure in only 7:37~min, roughly $44\%$ faster.

\paragraph*{Spin Correlations and MadSpin}

In $t\bar{t}$ production, the spins of the tops are correlated, which affects the angular distributions of the decay products. \mg can include these correlations through the decay chain syntax, decaying unstable particles with the Breit-Wigner propagator, or through MadSpin, a dedicated tool that decays particles after the hard process has been generated. We ask whether spin correlations are fully preserved in the decay chain syntax for leptonic decays, how this compares to MadSpin, and what reweighting formula MadSpin uses internally. This produces three distinct errors, each caused by a different problem in our curated documentation.
\begin{itemize}
    \item Concerning NLO compatibility, \ma state that the decay chain syntax works at NLO, but it does not. Our documentation did not mention this restriction. Adding a single warning sentence resolves the error.
    \item For the MadSpin reweighting formula, \ma present a reweighting formula that is correct for one of MadSpin's two operating modes, but not for the default mode. Adding a single sentence to our documentation, clarifying which mode the formula applies to, is sufficient.
    \item \ma confuse two standard labels used in the NLO matching formalism because our documentation buries the definitions in an unrelated section. Reformatting the existing text into a visible list resolves the error.
\end{itemize}

\paragraph*{} The \mg documentation enhances \ma' out-of-the-box performance, while the self-improvement loop shifts the focus from built-in knowledge to reasoning capabilities, and enables controlled, individualized adaptation of \ma.

\clearpage
\bibliography{tilman,refs}

@article{Janssen:2025zke,
    author = "Jan{\ss}en, Timo and Poncelet, Rene and Schumann, Steffen",
    title = "{Sampling NNLO QCD phase space with normalizing flows}",
    eprint = "2505.13608",
    archivePrefix = "arXiv",
    primaryClass = "hep-ph",
    reportNumber = "IFJPAN-IV-2025-11, MCNET-25-11, COMETA-2025-22",
    doi = "10.1007/JHEP09(2025)194",
    journal = "JHEP",
    volume = "09",
    pages = "194",
    year = "2025"
}

@article{Herrmann:2025nnz,
    author = "Herrmann, Tim and Jan{\ss}en, Timo and Schenker, Mathis and Schumann, Steffen and Siegert, Frank",
    title = "{Accelerating multijet-merged event generation with neural network matrix element surrogates}",
    eprint = "2506.06203",
    archivePrefix = "arXiv",
    primaryClass = "hep-ph",
    reportNumber = "MCNET-25-12",
    month = "6",
    year = "2025"
}

@article{Chan:2023icm,
    author = "Chan, Jay and Ju, Xiangyang and Kania, Adam and Nachman, Benjamin and Sangli, Vishnu and Siodmok, Andrzej",
    title = "{Integrating particle flavor into deep learning models for hadronization}",
    eprint = "2312.08453",
    archivePrefix = "arXiv",
    primaryClass = "hep-ph",
    doi = "10.1103/hgbg-k7js",
    journal = "Phys. Rev. D",
    volume = "111",
    number = "11",
    pages = "116015",
    year = "2025"
}

@article{Schwartz:2026ekw,
    author = "Schwartz, Matthew D.",
    title = "{Resummation of the C-Parameter Sudakov Shoulder Using Effective Field Theory}",
    eprint = "2601.02484",
    archivePrefix = "arXiv",
    primaryClass = "hep-ph",
    month = "1",
    year = "2026"
}

@article{Bewick:2023tfi,
    author = "Bewick, Gavin and others",
    title = "{Herwig 7.3 release note}",
    eprint = "2312.05175",
    archivePrefix = "arXiv",
    primaryClass = "hep-ph",
    reportNumber = "CERN-TH-2023-223, HERWIG-2023-01, KA-TP-28-2023, MCnet-23-19, IPPP/23/66",
    doi = "10.1140/epjc/s10052-024-13211-9",
    journal = "Eur. Phys. J. C",
    volume = "84",
    number = "10",
    pages = "1053",
    year = "2024"
}

@article{Sherpa:2019gpd,
    author = "Bothmann, Enrico and others",
    collaboration = "Sherpa",
    title = "{Event Generation with Sherpa 2.2}",
    eprint = "1905.09127",
    archivePrefix = "arXiv",
    primaryClass = "hep-ph",
    reportNumber = "FERMILAB-PUB-19-218-T, SLAC-PUB-17433, IPPP/19/42, MCNET-19-11",
    doi = "10.21468/SciPostPhys.7.3.034",
    journal = "SciPost Phys.",
    volume = "7",
    number = "3",
    pages = "034",
    year = "2019"
}

@article{Alwall:2014hca,
    author = "Alwall, J. and Frederix, R. and Frixione, S. and Hirschi, V. and Maltoni, F. and Mattelaer, O. and Shao, H. -S. and Stelzer, T. and Torrielli, P. and Zaro, M.",
    title = "{The automated computation of tree-level and next-to-leading order differential cross sections, and their matching to parton shower simulations}",
    eprint = "1405.0301",
    archivePrefix = "arXiv",
    primaryClass = "hep-ph",
    reportNumber = "CERN-PH-TH-2014-064, CP3-14-18, LPN14-066, MCNET-14-09, ZU-TH-14-14",
    doi = "10.1007/JHEP07(2014)079",
    journal = "JHEP",
    volume = "07",
    pages = "079",
    year = "2014"
}

@article{Sjostrand:2014zea,
    author = {Sj{\"o}strand, Torbj{\"o}rn and Ask, Stefan and Christiansen, Jesper R. and Corke, Richard and Desai, Nishita and Ilten, Philip and Mrenna, Stephen and Prestel, Stefan and Rasmussen, Christine O. and Skands, Peter Z.},
    title = "{An introduction to PYTHIA 8.2}",
    eprint = "1410.3012",
    archivePrefix = "arXiv",
    primaryClass = "hep-ph",
    reportNumber = "LU-TP-14-36, MCNET-14-22, CERN-PH-TH-2014-190, FERMILAB-PUB-14-316-CD, DESY-14-178, SLAC-PUB-16122",
    doi = "10.1016/j.cpc.2015.01.024",
    journal = "Comput. Phys. Commun.",
    volume = "191",
    pages = "159--177",
    year = "2015"
}

@article{Alwall:2011uj,
    author = "Alwall, Johan and Herquet, Michel and Maltoni, Fabio and Mattelaer, Olivier and Stelzer, Tim",
    title = "{MadGraph 5 : Going Beyond}",
    eprint = "1106.0522",
    archivePrefix = "arXiv",
    primaryClass = "hep-ph",
    reportNumber = "FERMILAB-PUB-11-448-T",
    doi = "10.1007/JHEP06(2011)128",
    journal = "JHEP",
    volume = "06",
    pages = "128",
    year = "2011"
}

@article{Maltoni:2002qb,
    author = "Maltoni, Fabio and Stelzer, Tim",
    title = "{MadEvent: Automatic event generation with MadGraph}",
    eprint = "hep-ph/0208156",
    archivePrefix = "arXiv",
    doi = "10.1088/1126-6708/2003/02/027",
    journal = "JHEP",
    volume = "02",
    pages = "027",
    year = "2003"
}

@article{ting2024astromlab1winsastronomy,
      title={AstroMLab 1: Who Wins Astronomy Jeopardy!?}, 
      author={Yuan-Sen Ting and Tuan Dung Nguyen and Tirthankar Ghosal and Rui Pan and Hardik Arora and Zechang Sun and Tijmen de Haan and Nesar Ramachandra and Azton Wells and Sandeep Madireddy and Alberto Accomazzi},
      year={2024},
      eprint={2407.11194},
      archivePrefix={arXiv},
      primaryClass={astro-ph.IM},
}

@article{UniverseTBD:2024pmh,
    author = "Perkowski, Ernest and others",
    collaboration = "UniverseTBD",
    title = "{AstroLLaMA-Chat: Scaling AstroLLaMA with Conversational and Diverse Datasets}",
    eprint = "2401.01916",
    archivePrefix = "arXiv",
    primaryClass = "astro-ph.IM",
    doi = "10.3847/2515-5172/ad1abe",
    journal = "Res. Notes AAS",
    volume = "8",
    number = "1",
    pages = "7",
    year = "2024"
}

@article{grezes2021buildingastrobertlanguagemodel,
      title={Building astroBERT, a language model for Astronomy \& Astrophysics}, 
      author={Felix Grezes and Sergi Blanco-Cuaresma and Alberto Accomazzi and Michael J. Kurtz and Golnaz Shapurian and Edwin Henneken and Carolyn S. Grant and Donna M. Thompson and Roman Chyla and Stephen McDonald and Timothy W. Hostetler and Matthew R. Templeton and Kelly E. Lockhart and Nemanja Martinovic and Shinyi Chen and Chris Tanner and Pavlos Protopapas},
      year={2021},
      eprint={2112.00590},
      archivePrefix={arXiv},
      primaryClass={cs.CL},
}

@article{Menzo:2025cim,
    author = {Menzo, Tony and Roman, Alexander and Gleyzer, Sergei and Matchev, Konstantin and Fleming, George T. and H{\"o}che, Stefan and Mrenna, Stephen and Shyamsundar, Prasanth},
    title = "{HEPTAPOD: Orchestrating High Energy Physics Workflows Towards Autonomous Agency}",
    eprint = "2512.15867",
    archivePrefix = "arXiv",
    primaryClass = "hep-ph",
    reportNumber = "FERMILAB-PUB-25-0923-CSAID-ETD-T",
    month = "12",
    year = "2025"
}

@article{Moss:2025ynt,
    author = "Moss, Adam",
    title = "{The AI Cosmologist I: An Agentic System for Automated Data Analysis}",
    eprint = "2504.03424",
    archivePrefix = "arXiv",
    primaryClass = "astro-ph.IM",
    month = "4",
    year = "2025"
}

@article{sun2025interpretingmultibandgalaxyobservations,
      title={Interpreting Multi-band Galaxy Observations with Large Language Model-Based Agents}, 
      author={Zechang Sun and Yuan-Sen Ting and Yaobo Liang and Nan Duan and Song Huang and Zheng Cai},
      year={2025},
      eprint={2409.14807},
      archivePrefix={arXiv},
      primaryClass={astro-ph.IM},
}

@article{deHaan:2024ybs,
    author = "de Haan, Tijmen",
    title = "{cosmosage: A natural-language assistant for cosmology}",
    eprint = "2407.04420",
    archivePrefix = "arXiv",
    primaryClass = "astro-ph.IM",
    doi = "10.1016/j.ascom.2025.100934",
    journal = "Astron. Comput.",
    volume = "51",
    pages = "100934",
    year = "2025"
}

@article{Nguyen:2023nhp,
    author = "Nguyen, Tuan Dung and others",
    title = "{AstroLLaMA: Towards Specialized Foundation Models in Astronomy}",
    eprint = "2309.06126",
    archivePrefix = "arXiv",
    primaryClass = "astro-ph.IM",
    month = "9",
    year = "2023"
}

@article{Laverick:2024fyh,
    author = "Laverick, Andrew and Surrao, Kristen and Zubeldia, Inigo and Bolliet, Boris and Cranmer, Miles and Lewis, Antony and Sherwin, Blake and Lesgourgues, Julien",
    title = "{Multi-Agent System for Cosmological Parameter Analysis}",
    eprint = "2412.00431",
    archivePrefix = "arXiv",
    primaryClass = "astro-ph.IM",
    month = "11",
    year = "2024"
}

@article{Zhang:2025uoo,
    author = "Zhang, Xiaowen and Bi, Zhenyu and Lachance, Patrick and Wang, Xuan and Di Matteo, Tiziana and Croft, Rupert A. C.",
    title = "{Bridging Literature and the Universe Via A Multi-Agent Large Language Model System}",
    eprint = "2507.08958",
    archivePrefix = "arXiv",
    primaryClass = "astro-ph.IM",
    month = "7",
    year = "2025"
}

@article{Richmond:2025lzg,
    author = "Richmond, Paul and Papageorgakis, Constantinos and Niarchos, Vasilis and Chowdhury, Borun and Agarwal, Prarit",
    title = "{FeynTune: large language models for high-energy theory}",
    eprint = "2508.03716",
    archivePrefix = "arXiv",
    primaryClass = "cs.CL",
    doi = "10.1088/2632-2153/ae47bb",
    journal = "Mach. Learn. Sci. Tech.",
    volume = "7",
    number = "2",
    pages = "025012",
    year = "2026"
}

@article{Bakshi:2025fgx,
    author = "Bakshi, S. D. and others",
    title = "{ArgoLOOM: agentic AI for fundamental physics from quarks to cosmos}",
    eprint = "2510.02426",
    archivePrefix = "arXiv",
    primaryClass = "hep-ph",
    reportNumber = "ANL-199516",
    month = "10",
    year = "2025"
}

@article{Gendreau-Distler:2025fsj,
    author = "Gendreau-Distler, Eli and Ho, Joshua and Kim, Dongwon and Le Pottier, Luc Tomas and Wang, Haichen and Yang, Chengxi",
    title = "{Automating High Energy Physics Data Analysis with LLM-Powered Agents}",
    booktitle = "{39th Annual Conference on Neural Information Processing Systems}: {Includes Machine Learning and the Physical Sciences (ML4PS)}",
    eprint = "2512.07785",
    archivePrefix = "arXiv",
    primaryClass = "physics.data-an",
    month = "12",
    year = "2025"
}

@article{Diefenbacher:2025zzn,
    author = {Diefenbacher, Sascha and Hallin, Anna and Kasieczka, Gregor and Kr{\"a}mer, Michael and Lauscher, Anne and Lukas, Tim},
    title = "{Agents of Discovery}",
    eprint = "2509.08535",
    archivePrefix = "arXiv",
    primaryClass = "hep-ph",
    month = "9",
    year = "2025"
}

@article{zhao2023survey,
  title={A survey of large language models},
  author={Zhao, Wayne Xin and Zhou, Kun and Li, Junyi and Tang, Tianyi and Wang, Xiaolei and Hou, Yupeng and Min, Yingqian and Zhang, Beichen and Zhang, Junjie and Dong, Zican and others},
  archivePrefix = "arXiv",
  eprint = "2303.18223",
  primaryClass = "cs.CL",
  volume={1},
  number={2},
  year={2023}
}

@article{yin2024survey,
  title={A survey on multimodal large language models},
  author={Yin, Shukang and Fu, Chaoyou and Zhao, Sirui and Li, Ke and Sun, Xing and Xu, Tong and Chen, Enhong},
  journal={National Science Review},
  volume={11},
  number={12},
  pages={nwae403},
  year={2024},
  publisher={Oxford University Press},
  archivePrefix = "arXiv",
  eprint = "2306.13549",
  primaryClass = "cs.CV",
}

@article{geng2023grammar,
  title={Grammar-constrained decoding for structured NLP tasks without finetuning},
  author={Geng, Saibo and Josifoski, Martin and Peyrard, Maxime and West, Robert},
  archivePrefix = "arXiv",
  eprint = "2305.13971",
  primaryClass = "cs.CL",
  year={2023}
}

@article{parisi2022talm,
  title={Talm: Tool augmented language models},
  author={Parisi, Aaron and Zhao, Yao and Fiedel, Noah},
  archivePrefix = "arXiv",
  eprint = "2205.12255",
  primaryClass = "cs.CL",
  year={2022}
}

@article{schick2023toolformer,
  title={Toolformer: Language models can teach themselves to use tools},
  author={Schick, Timo and Dwivedi-Yu, Jane and Dess{\`\i}, Roberto and Raileanu, Roberta and Lomeli, Maria and Hambro, Eric and Zettlemoyer, Luke and Cancedda, Nicola and Scialom, Thomas},
  journal={Advances in Neural Information Processing Systems},
  volume={36},
  pages={68539--68551},
  year={2023},
  archivePrefix = "arXiv",
  eprint = "2302.04761",
  primaryClass = "cs.CL",
}

@article{yao2022react,
  title={React: Synergizing reasoning and acting in language models},
  author={Yao, Shunyu and Zhao, Jeffrey and Yu, Dian and Du, Nan and Shafran, Izhak and Narasimhan, Karthik R and Cao, Yuan},
  booktitle={The eleventh international conference on learning representations},
  year={2022},
  archivePrefix = "arXiv",
  eprint = "2210.03629",
  primaryClass = "cs.CL",
}

@article{chen2023teaching,
  title={Teaching large language models to self-debug},
  author={Chen, Xinyun and Lin, Maxwell and Sch{\"a}rli, Nathanael and Zhou, Denny},
  year={2023},
  archivePrefix = "arXiv",
  eprint = "2304.05128",
  primaryClass = "cs.CL",
}

@article{wang2023voyager,
  title={Voyager: An open-ended embodied agent with large language models},
  author={Wang, Guanzhi and Xie, Yuqi and Jiang, Yunfan and Mandlekar, Ajay and Xiao, Chaowei and Zhu, Yuke and Fan, Linxi and Anandkumar, Anima},
  year={2023},
  archivePrefix = "arXiv",
  eprint = "2305.16291",
  primaryClass = "cs.AI",
}

@article{cai2023large,
  title={Large language models as tool makers},
  author={Cai, Tianle and Wang, Xuezhi and Ma, Tengyu and Chen, Xinyun and Zhou, Denny},
  year={2023},
  archivePrefix = "arXiv",
  eprint = "2305.17126",
  primaryClass = "cs.LG",
}

@article{yang2024swe,
  title={Swe-agent: Agent-computer interfaces enable automated software engineering},
  author={Yang, John and Jimenez, Carlos E and Wettig, Alexander and Lieret, Kilian and Yao, Shunyu and Narasimhan, Karthik and Press, Ofir},
  journal={Advances in Neural Information Processing Systems},
  volume={37},
  pages={50528--50652},
  year={2024},
  archivePrefix = "arXiv",
  eprint = "2405.15793",
  primaryClass = "cs.SE",
}

@article{wei2022chain,
  title={Chain-of-thought prompting elicits reasoning in large language models},
  author={Wei, Jason and Wang, Xuezhi and Schuurmans, Dale and Bosma, Maarten and Xia, Fei and Chi, Ed and Le, Quoc V and Zhou, Denny and others},
  journal={Advances in neural information processing systems},
  volume={35},
  pages={24824--24837},
  year={2022},
  archivePrefix = "arXiv",
  eprint = "2201.11903",
  primaryClass = "cs.CL",
}

@article{du2023improving,
  title={Improving factuality and reasoning in language models through multiagent debate},
  author={Du, Yilun and Li, Shuang and Torralba, Antonio and Tenenbaum, Joshua B and Mordatch, Igor},
  booktitle={Forty-first International Conference on Machine Learning},
  year={2023},
  archivePrefix = "arXiv",
  eprint = "2305.14325",
  primaryClass = "cs.CL",
}

@article{liu2023dynamic,
  title={Dynamic llm-agent network: An llm-agent collaboration framework with agent team optimization},
  author={Liu, Zijun and Zhang, Yanzhe and Li, Peng and Liu, Yang and Yang, Diyi},
  year={2023},
  archivePrefix = "arXiv",
  eprint = "2310.02170",
  primaryClass = "cs.CL",
}

@article{hong2023metagpt,
  title={MetaGPT: Meta programming for a multi-agent collaborative framework},
  author={Hong, Sirui and Zhuge, Mingchen and Chen, Jonathan and Zheng, Xiawu and Cheng, Yuheng and Wang, Jinlin and Zhang, Ceyao and Wang, Zili and Yau, Steven Ka Shing and Lin, Zijuan and others},
  booktitle={The Twelfth International Conference on Learning Representations},
  year={2023},
  archivePrefix = "arXiv",
  eprint = "2308.00352",
  primaryClass = "cs.AI",
}

@article{wang2024survey,
  title={A survey on large language model based autonomous agents},
  author={Wang, Lei and Ma, Chen and Feng, Xueyang and Zhang, Zeyu and Yang, Hao and Zhang, Jingsen and Chen, Zhiyuan and Tang, Jiakai and Chen, Xu and Lin, Yankai and others},
  journal={Frontiers of Computer Science},
  volume={18},
  number={6},
  pages={186345},
  year={2024},
  publisher={Springer},
  archivePrefix = "arXiv",
  eprint = "2308.11432",
  primaryClass = "cs.AI",
}

@article{chen2024survey,
  title={A survey on llm-based multi-agent system: Recent advances and new frontiers in application},
  author={Chen, Shuaihang and Liu, Yuanxing and Han, Wei and Zhang, Weinan and Liu, Ting},
  year={2024},
  archivePrefix = "arXiv",
  eprint = "2412.17481",
  primaryClass = "cs.CL",
}

@article{Alloul:2013bka,
  author         = "Alloul, Adam and Christensen, Neil D. and Degrande, Celine and Duhr, Claude and Fuks, Benjamin",
  title          = "FeynRules 2.0 - A complete toolbox for tree-level phenomenology",
  journal        = "Comput. Phys. Commun.",
  volume         = "185",
  year           = "2014",
  pages          = "2250--2300",
  eprint         = "1310.1921",
  archivePrefix  = "arXiv",
  primaryClass   = "hep-ph"
}

@article{Ball:2017nwa,
  author         = "Ball, Richard D. and others",
  title          = "Parton distributions from high-precision collider data",
  journal        = "Eur. Phys. J. C",
  volume         = "77",
  year           = "2017",
  pages          = "663",
  eprint         = "1706.00428",
  archivePrefix  = "arXiv",
  primaryClass   = "hep-ph"
}

@article{Dobbs:2001ck,
  author         = "Dobbs, Matt and Hansen, John B.",
  title          = "The HepMC C++ Monte Carlo event record for High Energy Physics",
  journal        = "Comput. Phys. Commun.",
  volume         = "134",
  year           = "2001",
  pages          = "41--46",
  eprint         = "hep-ph/0011365",
  archivePrefix  = "arXiv",
  primaryClass   = "hep-ph"
}

@article{Dorsner:2016wpm,
  author         = "Dorsner, Ilja and Greljo, Admir and Kamenik, Jernej F. and Kosnik, Nejc",
  title          = "Physics of leptoquarks in precision experiments and at particle colliders",
  journal        = "Phys. Rept.",
  volume         = "641",
  year           = "2016",
  pages          = "1--68",
  eprint         = "1603.04993",
  archivePrefix  = "arXiv",
  primaryClass   = "hep-ph"
}

@article{Cacciari:2011ma,
  author         = "Cacciari, Matteo and Salam, Gavin P. and Soyez, Gregory",
  title          = "FastJet User Manual",
  journal        = "Eur. Phys. J. C",
  volume         = "72",
  year           = "2012",
  pages          = "1896",
  eprint         = "1111.6097",
  archivePrefix  = "arXiv",
  primaryClass   = "hep-ph"
}

@article{Cacciari:2008gp,
  author         = "Cacciari, Matteo and Salam, Gavin P. and Soyez, Gregory",
  title          = "The anti-$k_t$ jet clustering algorithm",
  journal        = "JHEP",
  volume         = "04",
  year           = "2008",
  pages          = "063",
  eprint         = "0802.1189",
  archivePrefix  = "arXiv",
  primaryClass   = "hep-ph"
}

@article{Buckley:2014ana,
    author = {Buckley, Andy and Ferrando, James and Lloyd, Stephen and Nordstr{\"o}m, Karl and Page, Ben and R{\"u}fenacht, Martin and Sch{\"o}nherr, Marek and Watt, Graeme},
    title = "{LHAPDF6: parton density access in the LHC precision era}",
    eprint = "1412.7420",
    archivePrefix = "arXiv",
    primaryClass = "hep-ph",
    reportNumber = "GLAS-PPE-2014-05, MCNET-14-29, IPPP-14-111, DCPT-14-222",
    doi = "10.1140/epjc/s10052-015-3318-8",
    journal = "Eur. Phys. J. C",
    volume = "75",
    pages = "132",
    year = "2015"
}

@article{Artoisenet:2012st,
    author = "Artoisenet, Pierre and Frederix, Rikkert and Mattelaer, Olivier and Rietkerk, Robbert",
    title = "{Automatic spin-entangled decays of heavy resonances in Monte Carlo simulations}",
    eprint = "1212.3460",
    archivePrefix = "arXiv",
    primaryClass = "hep-ph",
    reportNumber = "NIKHEF-2012-021, CERN-PH-TH-2012-329",
    doi = "10.1007/JHEP03(2013)015",
    journal = "JHEP",
    volume = "03",
    pages = "015",
    year = "2013"
}

@article{mukund2026marvel,
  title={MARVEL: A Multi Agent-based Research Validator and Enabler using Large Language Models},
  author={Mukund, Nikhil and Luo, Yifang and Zhang, Fan and Barsotti, Lisa and Katsavounidis, Erik},
  year={2026},
  archivePrefix = "arXiv",
  eprint = "2601.03436",
  primaryClass = "astro-ph.IM",
}

@article{nye2021show,
  title={Show your work: Scratchpads for intermediate computation with language models},
  author={Nye, Maxwell and Andreassen, Anders Johan and Gur-Ari, Guy and Michalewski, Henryk and Austin, Jacob and Bieber, David and Dohan, David and Lewkowycz, Aitor and Bosma, Maarten and Luan, David and others},
  year={2021},
  archivePrefix = "arXiv",
  eprint = "2112.00114",
  primaryClass = "cs.LG",
}

@inproceedings{song2023llm,
  title={Llm-planner: Few-shot grounded planning for embodied agents with large language models},
  author={Song, Chan Hee and Wu, Jiaman and Washington, Clayton and Sadler, Brian M and Chao, Wei-Lun and Su, Yu},
  booktitle={Proceedings of the IEEE/CVF international conference on computer vision},
  pages={2998--3009},
  year={2023},
  archivePrefix = "arXiv",
  eprint = "2212.04088",
  primaryClass = "cs.AI",
}

@article{d2024marg,
  title={Marg: Multi-agent review generation for scientific papers},
  author={D'Arcy, Mike and Hope, Tom and Birnbaum, Larry and Downey, Doug},
  year={2024},
  archivePrefix = "arXiv",
  eprint = "2401.04259",
  primaryClass = "cs.CL",
}

@article{shinn2023reflexion,
  title={Reflexion: Language agents with verbal reinforcement learning},
  author={Shinn, Noah and Cassano, Federico and Gopinath, Ashwin and Narasimhan, Karthik and Yao, Shunyu},
  journal={Advances in neural information processing systems},
  volume={36},
  pages={8634--8652},
  year={2023},
  archivePrefix = "arXiv",
  eprint = "2303.11366",
  primaryClass = "cs.AI",
}

@inproceedings{zhao2024expel,
  title={Expel: Llm agents are experiential learners},
  author={Zhao, Andrew and Huang, Daniel and Xu, Quentin and Lin, Matthieu and Liu, Yong-Jin and Huang, Gao},
  booktitle={Proceedings of the AAAI Conference on Artificial Intelligence},
  volume={38},
  number={17},
  pages={19632--19642},
  year={2024},
  archivePrefix = "arXiv",
  eprint = "2308.10144",
  primaryClass = "cs.LG",
}

@article{Bahl:2026qaf,
    author = "Bahl, Henning and Braun, Jens and Heinrich, Gudrun and Plehn, Tilman and Revelli, Rebecca",
    title = "{How to Trust Learned Loop Amplitudes}",
    eprint = "2601.00950",
    archivePrefix = "arXiv",
    primaryClass = "hep-ph",
    month = "1",
    year = "2026"
}

@article{Butter:2025wxn,
    author = "Butter, Anja and others",
    title = "{Iterative HOMER with uncertainties}",
    eprint = "2509.03592",
    archivePrefix = "arXiv",
    primaryClass = "hep-ph",
    reportNumber = "FERMILAB-PUB-25-0579-CSAID",
    month = "9",
    year = "2025"
}

@article{Bahl:2025xvx,
    author = "Bahl, Henning and Elmer, Nina and Plehn, Tilman and Winterhalder, Ramon",
    title = "{Amplitude Uncertainties Everywhere All at Once}",
    eprint = "2509.00155",
    archivePrefix = "arXiv",
    primaryClass = "hep-ph",
    reportNumber = "TIF-UNIMI-2025-17",
    month = "8",
    year = "2025"
}

@article{Heneka:2025fpe,
    author = "Heneka, Caroline and Nieser, Florian and Ore, Ayodele and Plehn, Tilman and Schiller, Daniel",
    title = "{Large Language Models -- the Future of Fundamental Physics?}",
    eprint = "2506.14757",
    archivePrefix = "arXiv",
    primaryClass = "astro-ph.CO",
    month = "6",
    year = "2025"
}

@article{Bahl:2024gyt,
    author = "Bahl, Henning and Elmer, Nina and Favaro, Luigi and Hau{\ss}mann, Manuel and Plehn, Tilman and Winterhalder, Ramon",
    title = "{Accurate Surrogate Amplitudes with Calibrated Uncertainties}",
    eprint = "2412.12069",
    archivePrefix = "arXiv",
    primaryClass = "hep-ph",
    doi = "10.21468/SciPostPhysCore.8.4.073",
    journal = "SciPost Phys. Core",
    volume = "8",
    pages = "073",
    year = "2025"
}

@article{Heimel:2024wph,
    author = "Heimel, Theo and Mattelaer, Olivier and Plehn, Tilman and Winterhalder, Ramon",
    title = "{Differentiable MadNIS-Lite}",
    eprint = "2408.01486",
    archivePrefix = "arXiv",
    primaryClass = "hep-ph",
    reportNumber = "IRMP-CP3-24-23",
    doi = "10.21468/SciPostPhys.18.1.017",
    journal = "SciPost Phys.",
    volume = "18",
    number = "1",
    pages = "017",
    year = "2025"
}

@article{Elmer:2023wtr,
    author = "Elmer, Nina and Madigan, Maeve and Plehn, Tilman and Schmal, Nikita",
    title = "{Staying on Top of SMEFT-Likelihood Analyses}",
    eprint = "2312.12502",
    archivePrefix = "arXiv",
    primaryClass = "hep-ph",
    doi = "10.21468/SciPostPhys.18.3.108",
    journal = "SciPost Phys.",
    volume = "18",
    pages = "108",
    year = "2025"
}

@article{Heimel:2023ngj,
    author = "Heimel, Theo and Huetsch, Nathan and Maltoni, Fabio and Mattelaer, Olivier and Plehn, Tilman and Winterhalder, Ramon",
    title = "{The MadNIS reloaded}",
    eprint = "2311.01548",
    archivePrefix = "arXiv",
    primaryClass = "hep-ph",
    reportNumber = "IRMP-CP3-23-56, MCNET-23-12",
    doi = "10.21468/SciPostPhys.17.1.023",
    journal = "SciPost Phys.",
    volume = "17",
    number = "1",
    pages = "023",
    year = "2024"
}

@article{Heimel:2022wyj,
    author = "Heimel, Theo and Winterhalder, Ramon and Butter, Anja and Isaacson, Joshua and Krause, Claudius and Maltoni, Fabio and Mattelaer, Olivier and Plehn, Tilman",
    title = "{MadNIS - Neural multi-channel importance sampling}",
    eprint = "2212.06172",
    archivePrefix = "arXiv",
    primaryClass = "hep-ph",
    reportNumber = "IRMP-CP3-22-56, MCNET-22-22, FERMILAB-PUB-22-915-T",
    doi = "10.21468/SciPostPhys.15.4.141",
    journal = "SciPost Phys.",
    volume = "15",
    number = "4",
    pages = "141",
    year = "2023"
}

@article{Plehn:2022ftl,
    author = "Plehn, Tilman and Butter, Anja and Dillon, Barry and Heimel, Theo and Krause, Claudius and Winterhalder, Ramon",
    title = "{Modern Machine Learning for LHC Physicists}",
    eprint = "2211.01421",
    archivePrefix = "arXiv",
    primaryClass = "hep-ph",
    month = "11",
    year = "2022"
}

@article{Butter:2022rso,
    author = "Badger, Simon and others",
    editor = "Butter, Anja and Plehn, Tilman and Schumann, Steffen",
    title = "{Machine learning and LHC event generation}",
    eprint = "2203.07460",
    archivePrefix = "arXiv",
    primaryClass = "hep-ph",
    reportNumber = "FERMILAB-PUB-22-126-T",
    doi = "10.21468/SciPostPhys.14.4.079",
    journal = "SciPost Phys.",
    volume = "14",
    number = "4",
    pages = "079",
    year = "2023"
}

@article{Alwall:2007st,
    author = "Alwall, Johan and Demin, Pavel and de Visscher, Simon and Frederix, Rikkert and Herquet, Michel and Maltoni, Fabio and Plehn, Tilman and Rainwater, David L. and Stelzer, Tim",
    title = "{MadGraph/MadEvent v4: The New Web Generation}",
    eprint = "0706.2334",
    archivePrefix = "arXiv",
    primaryClass = "hep-ph",
    reportNumber = "SLAC-PUB-12603, CP3-07-17",
    doi = "10.1088/1126-6708/2007/09/028",
    journal = "JHEP",
    volume = "09",
    pages = "028",
    year = "2007"
}

\end{document}